\documentclass[12pt]{article}
\pdfoutput=1

\usepackage{amsmath,amssymb,amsfonts,graphicx,amsfonts}
\usepackage{textcomp,mathcomp}
\usepackage[svgnames]{xcolor}
\usepackage[colorlinks,citecolor=DarkGreen,linkcolor=FireBrick]{hyperref}
\usepackage[noabbrev]{cleveref}
\usepackage{ascmac}
\usepackage{enumitem}
\usepackage{comment} 
\usepackage[skip=20pt]{subcaption}
\usepackage[braket, qm]{qcircuit}
\newcommand{\ii}{i}
\newcommand{\dd}{\operatorname{d}}
\newcommand{\col}{\operatorname{col}}
\newcommand{\Tr}{\operatorname{Tr}}

\newcommand{\kinetic}{\operatorname{K}}
\renewcommand{\vec}{\mathbf}

\allowdisplaybreaks[4]

\begin{document}
\begin{center}
{\Large A model of randomly-coupled Pauli spins}\\

\end{center}
\vspace{0.1cm}
\vspace{0.1cm}
\begin{center}

Masanori Hanada$^a$, Antal Jevicki$^b$, Xianlong Liu$^b$, Enrico Rinaldi$^c$, Masaki Tezuka$^d$\\

\end{center}
\vspace{0.3cm}

\begin{center}
$^a$School of Mathematical Sciences, Queen Mary University of London\\
Mile End Road, London, E1 4NS, United Kingdom\\
\vspace{3mm}
$^{a}$qBraid Co., Harper Court 5235, Chicago, IL 60615, USA\\
\vspace{3mm}
$^b$Department of Physics, Brown University\\
182 Hope Street, Providence, RI 02912, United States\\
\vspace{3mm}
$^b$Brown Theoretical Physics Center, Brown University, \\ 
340 Brook Street, Providence, RI 02912, United States\\
\vspace{3mm}
$^c$Quantinuum K.K., Otemachi Financial City Grand Cube 3F, 1-9-2 Otemachi,\\
Chiyoda-ku, Tokyo, Japan\\
\vspace{3mm}
$^c$Center for Quantum Computing (RQC), RIKEN\\
Wako, Saitama 351-0198, Japan\\
\vspace{3mm}
$^c$Theoretical  Quantum  Physics  Laboratory, RIKEN,
Wako, Saitama 351-0198, Japan\\
\vspace{3mm}
$^c$Interdisciplinary Theoretical and Mathematical Sciences (iTHEMS) Program, RIKEN\\
Wako, Saitama 351-0198, Japan\\
\vspace{3mm}
$^d$Department of Physics, Kyoto University\\
Kitashirakawa, Sakyo-ku, Kyoto 606-8502, Japan
\end{center}

\vspace{1.5cm}

\begin{center}
  {\bf Abstract}
\end{center}
We construct a model of Pauli spin operators with all-to-all $4$-local interactions by replacing Majorana fermions in the SYK model with spin operators. Equivalently, we replace fermions with hard-core bosons. We study this model numerically and compare the properties with those of the SYK model. We observe a striking quantitative coincidence between the spin model and the SYK model, which suggests that this spin model is strongly chaotic and, perhaps, can play some role in holography. We also discuss the path-integral approach with multi-local fields and the possibility of quantum simulations. This model may be an interesting target for quantum simulations because Pauli spins are easier to implement than fermions on qubit-based quantum devices.

\tableofcontents
\section{Introduction}
The Sachdev-Ye-Kitaev (SYK) model has been intensively studied for various motivations, ranging from condensed matter physics to quantum gravity via holography. Given the importance of the SYK model, it is natural to try quantum simulations. Indeed, there are a few attempts~\cite{Luo:2017bno,Jafferis:2022crx}. Still, it is difficult to simulate the SYK model on quantum devices without some simplifications. One of the obstacles is that elementary degrees of freedom are fermions and fermions are non local when mapped to qubits. Specifically, via the Jordan-Wigner transform, Majorana fermions $\hat{\chi}_a$ ($a=1,2,\cdots,N_{\rm Maj}$) satisfying $\{\hat{\chi}_a,\hat{\chi}_b\}=2\delta_{ab}$ are written in terms of Pauli strings (tensor products of Pauli matrices) acting on $N_{\rm spin}=\frac{N_{\rm Maj}}{2}$ spins as
\begin{align}
\hat{\chi}_1
&=
\hat{\sigma}_x\otimes\hat{I}\otimes\hat{I}\otimes\cdots\otimes\hat{I}\otimes\hat{I}\, , 
\nonumber\\
\hat{\chi}_2
&=
\hat{\sigma}_y\otimes\hat{I}\otimes\hat{I}\otimes\cdots\otimes\hat{I}\otimes\hat{I}\, , 
\nonumber\\
\hat{\chi}_3
&=
\hat{\sigma}_z\otimes\hat{\sigma}_x\otimes\hat{I}\otimes\cdots\otimes\hat{I}\otimes\hat{I}\, , 
\nonumber\\
\hat{\chi}_4
&=
\hat{\sigma}_z\otimes\hat{\sigma}_y\otimes\hat{I}\otimes\cdots\otimes\hat{I}\otimes\hat{I}\, , 
\nonumber\\
&
\qquad\qquad\quad\cdots
\nonumber\\
\hat{\chi}_{2N_{\rm spin}-1}
&=
\hat{\sigma}_z\otimes\hat{\sigma}_z\otimes\hat{\sigma}_z\otimes\cdots\otimes\hat{\sigma}_z\otimes\hat{\sigma}_x\, , 
\nonumber\\
\hat{\chi}_{2N_{\rm spin}}
&=
\hat{\sigma}_z\otimes\hat{\sigma}_z\otimes\hat{\sigma}_z\otimes\cdots\otimes\hat{\sigma}_z\otimes\hat{\sigma}_y\, .
\label{Jordan-Wigner}
\end{align}
Here, we used Pauli matrices
\begin{align}
    \hat{\sigma}_x=\begin{pmatrix}0&1\\1&0\end{pmatrix}\, ,
    \hat{\sigma}_y=\begin{pmatrix}0&-i\\i&0\end{pmatrix}\, ,
    \hat{\sigma}_z=\begin{pmatrix}1&0\\0&-1\end{pmatrix}\,   
\end{align}
and the identity matrix
\begin{align}
    \hat{I}=\begin{pmatrix}1&0\\0&1\end{pmatrix}, 
\end{align}
which act on the local Hilbert space of each qubit. These long chains of Pauli matrices, which are as long as the number of degrees of freedom in the system, require a lot of resources (quantum operations) in digital quantum simulations. 

In this paper, we will consider a spin model which is obtained by replacing all Pauli $\hat{\sigma}_z$ operators in \eqref{Jordan-Wigner} with the identity $\hat{I}$s. Such a theory contains only SU(2) spin variables (Pauli matrices $\hat{\sigma}_x$ and $\hat{\sigma}_y$) on $N_{\rm spin}$ sites. For brevity, we will denote this model as `SpinXY4' in this paper. $XY$ refers to $\hat{\sigma}_x$ and $\hat{\sigma}_y$ and 4 refers to the number of Pauli operators in the interaction.

There are several reasons we are interested in such a model. First of all, this model can be studied more easily on quantum computers. Therefore, if this model inherits some interesting features of the SYK model, it will be an interesting target for quantum simulation in the near future and, hopefully, serve as a good starting point for the experimental study of quantum gravity via holography~\cite{Danshita:2016xbo,Gharibyan:2020bab,Maldacena:2023acv}. 
We could hope that much of the physics is preserved by the replacement of fermions with spins, given that the Sachdev-Ye model (SY model)~\cite{Sachdev:1992fk}, which is closely related to the SYK model, is a model consisting of SU($M$) spin variables. A potential advantage of this model over the SY model is that there is only one limit ($N_{\rm spin}\to\infty$) while the SY model requires the large-spin limit ($M\to\infty$) and the many-spin limit ($N_{\rm spin}\to\infty$). The simple structure in terms of spin-$1/2$ variables makes the simulation on qubit-based quantum devices straightforward.\footnote{
Depending on the context, one could think the existence of two size parameters is the advantage that leads us to a richer phase diagram. Here, we regard the simplicity of the model as an advantage.
} Note that an important motivation for the large-$M$ limit in the SY model is to avoid the spin-glass phase, and hence, we would like to know if a spin-glass phase appears in SpinXY4.   

Our findings are the following:
\begin{itemize}
    \item 
    We studied the density of states (DoS) up to $N_{\rm Maj}=2N_{\rm spin}=34$ by exact numerical diagonalization, collecting many samples with different random couplings. For small $N_{\rm spin}$, the DoS is almost indistinguishable from the one for the SYK model. As $N_{\rm spin}$ increases, we see a small discrepancy near the edge, although the bulk of the spectrum looks very similar to SYK. (Sec.~\ref{sec:DoS})

    \item
    Statistical properties of the energy spectrum are consistent with those of Random Matrix Theory (RMT), suggesting the absence of the spin-glass phase except for a few low-energy modes. (Sec.~\ref{sec:Comparison_with_RMT})

    \item 
    The spectral form factor (SFF) has a long ramp that suggests a strongly-chaotic nature, similar to the SYK model. (Sec.~\ref{sec:Comparison_with_RMT})

    \item 
    For some values of $N_{\rm spin}$, some correlation functions are quantitatively close to the counterparts in SYK at any time scale. (Sec.~\ref{sec:SFF-all-time scale} and Sec.~\ref{sec:2-pt-fnc})

    \item
    While the Edwards-Anderson (EA) parameter defined using the $\hat{\sigma}_z$ operators as well as a generalized version of the EA parameter decrease monotonically as a function of the system size for a majority of the energy spectrum, their increase suggests that a small number of low-energy states behave as in the spin-glass state.
    (Sec.~\ref{sec:EA})
    
\end{itemize}
We believe these findings provide us with good motivation for further investigations. 

This paper is organized as follows. 
In Sec.~\ref{sec:definition} we give the precise definition of the model. We also provide an incomplete list of potential generalizations.
In Sec.~\ref{sec:DoS} we study the density of states. We make a quantitative comparison with the SYK model and find an intriguing resemblance, except for the edges. 
In Sec.~\ref{sec:Comparison_with_RMT} we study the correlation of energy eigenvalues and compare it with that of Random Matrix Theory. We observe striking similarities with the SYK model: agreement with RMT is observed except for a small number of low-lying modes, and the agreement extends to a wide energy band (equivalently, a long ramp is observed in the spectral form factor). 
In Sec.~\ref{sec:2-pt-fnc} we study two-point functions. The late-time behavior is consistent with RMT and similarities with the SYK model are observed. For certain choices of operators and values of $N$, we observe a striking quantitative coincidence at all time scales.
In Sec.~\ref{sec:EA}, we introduce a generalized version of the EA parameter, defined between eigenstates belonging to the two parity sectors, and study it along with the EA parameter.
In Sec.~\ref{sec:2-pt-fnc}, we study the two-point functions as a function of time and find strong similarities with the SYK model.
In Sec.~\ref{sec:path-integral} we introduce a path-integral formulation for the description of the large-$N$ limit of these spin models based on collective multi-local fields following a closed set of Schwinger-Dyson equations.
We end the paper by commenting on the implementation of a Trotterized Hamiltonian evolution of the SpinXY4 model on a quantum device in Sec.~\ref{sec:quantum-circuits} and then we conclude with an outlook.

\textit{Note added:} We note that SYK-like behavior has been observed in the spectral function of random Heisenberg magnets for low and finite frequencies \cite{PhysRevLett.126.136602}, where the ground state is spin glass \cite{PhysRevB.65.224430}. Also see \cite{swingle2023bosonic} for saddle-point equation study and numerical study for the SpinXYZ$q$ model (see below for definition), which was initially introduced as the quantum $p$-spin glass model~\cite{Erd_s_2014}.
Possibility of studying black hole spacetimes by a spin system with another type of four-spin random couplings has been previously discussed in \cite{1908.11190, 1912.04864}.

\section{Definition of the model}\label{sec:definition}
We use $N_\mathrm{spin}=\frac{N_{\rm Maj}}{2}$ spins instead of $N_{\rm Maj}$ Majorana fermions. 
Let $\hat{O}_{a}$ be the counterpart of $\hat{\chi}_{a}$, i.e., $\hat{O}_{a}$ is obtained by replacing $\hat{\sigma}_z$ with $\hat{I}$ in $\hat{\chi}_{a}$. Specifically, $\hat{O}_{2j-1}=\hat{\sigma}_{j,x}$ and $\hat{O}_{2j}=\hat{\sigma}_{j,y}$, where
\begin{align}
\hat{O}_1
&=
\hat{\sigma}_{1,x}
=
\hat{\sigma}_x\otimes\hat{I}\otimes\hat{I}\otimes\cdots\otimes\hat{I}\otimes\hat{I}\, , 
\nonumber\\
\hat{O}_2
&=
\hat{\sigma}_{1,y}
=
\hat{\sigma}_y\otimes\hat{I}\otimes\hat{I}\otimes\cdots\otimes\hat{I}\otimes\hat{I}\, , 
\nonumber\\
\hat{O}_3
&=
\hat{\sigma}_{2,x}
=
\hat{I}\otimes\hat{\sigma}_x\otimes\hat{I}\otimes\cdots\otimes\hat{I}\otimes\hat{I}\, , 
\nonumber\\
\hat{O}_4
&=
\hat{\sigma}_{2,y}
=
\hat{I}\otimes\hat{\sigma}_y\otimes\hat{I}\otimes\cdots\otimes\hat{I}\otimes\hat{I}\, , 
\nonumber\\
&
\qquad\qquad\quad\cdots
\nonumber\\
\hat{O}_{2N_{\rm spin}-1}
&=
\hat{\sigma}_{N_{\rm spin},x}
=
\hat{I}\otimes\hat{I}\otimes\hat{I}\otimes\cdots\otimes\hat{I}\otimes\hat{\sigma}_x\, , 
\nonumber\\
\hat{O}_{2N_{\rm spin}}
&=
\hat{\sigma}_{N_{\rm spin},y}
=
\hat{I}\otimes\hat{I}\otimes\hat{I}\otimes\cdots\otimes\hat{I}\otimes\hat{\sigma}_y\, .
\end{align}
The Hamiltonian of the model is the following: 
\begin{align}
    \hat{H}=\sqrt{\frac{6}{N_{\rm Maj}^3}}\sum_{1\le a<b<c<d\le N_{\rm Maj}}J_{abcd}i^{\eta_{abcd}}\hat{O}_a\hat{O}_b\hat{O}_c\hat{O}_d\, ,
    \label{spin_Hamiltonian}
\end{align}
in which the couplings $J_{abcd}$ are chosen from the standard normal distribution
\begin{align}
    P(J_{abcd})=\frac{1}{\sqrt{2\pi}}e^{-J_{abcd}^2/2},
\end{align}
and $\eta_{abcd}$ is the number of spins whose both $x$ and $y$ components appear in $(a,b,c,d)$, e.g., $\eta_{1357}=0$, $\eta_{1235}=1$, $\eta_{1234}=2$.
We need $i^{\eta_{abcd}}$ for the Hermiticity of the Hamiltonian.~\footnote{Note that $\hat{O}_{2j-1}\hat{O}_{2j}=\hat{\sigma}_{j,x}\hat{\sigma}_{j,y}
=i\hat{\sigma}_{j,z}$ is anti-Hermitian.}
We will compare this model with the SYK model with $q=4$ which we rename for conciseness as `SYK4':
\begin{align}
    \hat{H}_{\rm SYK}=\sqrt{\frac{6}{N_{\rm Maj}^3}}\sum_{1\leq a<b<c<d\leq N_{\rm Maj}}J_{abcd}\hat{\chi}_a\hat{\chi}_b\hat{\chi}_c\hat{\chi }_d\, . 
    \label{SYK_Hamiltonian}
\end{align}

We chose the normalization of the random couplings $J_{abcd}$ in such a way that the large-$N_{\rm Maj}$ limit of the SYK model simplifies. Specifically, the energy $E$ and entropy $S$ scale as $N_{\rm Maj}^1$ when the temperature $T$ is fixed to be an order-$N_{\rm Maj}^0$ value and characteristic time scales, such as the decay rate of a two-point function, are of order $N_{\rm Maj}^0$. 

Despite an apparent similarity at a formal level, the Hamiltonians \eqref{spin_Hamiltonian} and \eqref{SYK_Hamiltonian} are clearly different because we are using different building blocks: Pauli spins $\hat{O}$ in the former and fermions $\hat{\chi}$ in the latter. We could interpret $\hat{O}_{2a-1}\pm i\hat{O}_{2a}$ as the creation and annihilation operators of a hard-core boson\footnote{
A hard-core boson is bosonic in that there is no sign factor associated with the exchange of two of them but, unlike the usual bosons, only one hard-core boson can sit at each site. 
} rather than a fermion.

\subsection{Parity}
A convenient basis of the Hilbert space is $\{\ket{s_1,s_2,\cdots,s_{N_{\rm spin}}}\}$, where $s_a=\pm 1$ $(a=1,2,\cdots,N_{\rm spin})$ represents a spin up or spin down at each site. Because $\sigma_{a,x}$ and $\sigma_{a,y}$ change $s_a$ to $-s_a$ (up to down), and because the Hamiltonian is a sum of products of four of them, the product $\prod_{a=1}^{N_{\rm spin}}s_a$ is conserved. 
We can see this also by noticing that $\hat{H}$ and $\hat{\Gamma}\equiv\hat{\sigma}_{1,z}\otimes\hat{\sigma}_{2,z}\otimes\cdots\otimes\hat{\sigma}_{N_{\rm spin},z}$ commute. 
Therefore, the Hamiltonian can be written in a block-diagonal form with two blocks consisting of $\gamma\equiv\prod_{a=1}^{N_{\rm spin}}s_a=\pm 1$. Each block is a $2^{N_{\rm spin}-1}\times 2^{N_{\rm spin}-1}$ matrix. 

We will call $\gamma=+1$ the parity even sector and $\gamma=-1$ the parity odd sector. They correspond to the parity-even and odd sectors in the SYK model.

\subsection{Possible variants}\label{sec:variants}
Similarly to the case of the SYK model, we can consider many variants of the SpinXY4 model. 
\subsubsection*{$q$-local models (SpinXY$q$)}
We can take the number of spins in each interaction term to be a generic number $q$:
\begin{align}
    \hat{H}=\mathcal{N}\sum_{a_1<a_2<\cdots<a_q}J_{a_1a_2\cdots a_q}i^{\eta_{a_1a_2\cdots a_q}}\hat{O}_{a_1}\hat{O}_{a_2}\cdots\hat{O}_{a_q}\, .
\end{align}
where the standard choice of the normalization factor is $\mathcal{N}=\sqrt{q!(N_{\rm Maj}-q)!/N_{\rm Maj}!}$. 
\footnote{Unlike in the SYK case where the $q=2$ model is readily solvable, the $q=2$ case for the spin model is already nontrivial.}
Note that $q$ can be odd, in which case parity is not conserved.\footnote{
For the SYK model, it is not common to consider odd values of $q$ because the Hamiltonian would be fermionic then. Still, with an explicit basis choice, the Hamiltonian can be expressed as the ordinary Hermitian matrix and there is no apparent reason to exclude such models~\cite{Balasubramanian:2021mxo}.
}
\subsubsection*{Binary/Sparse model}
The random couplings $J_{abcd}$ can be made sparse (i.e., many of them can be set to be zero)~\cite{Xu:2020shn,Caceres:2021nsa,Caceres:2023yoj} and/or binary (nonzero couplings $\propto\pm1$)~\cite{Tezuka:2022mrr}.
\subsubsection*{Adding or removing $\sigma_z$}
In the SpinXY4 model defined above we allowed $\sigma_x$ and $\sigma_y$ on the same site to appear in the same interaction term. We can forbid this to happen and this amounts to setting $\eta=0$. Such a modification should not change the theory in the limit of $N_{\rm spin}\to\infty$. However, there are some differences that are not captured in the large-$N_{\rm spin}$ limit. For example, the universality class (from RMT) is the Gaussian unitary ensemble (GUE) for any $N_{\rm spin}$ when $\sigma_x$ and $\sigma_y$ are allowed at the same site, while it is the Gaussian orthogonal ensemble (GOE) for even $N_{\rm spin}$ and GUE for odd $N_{\rm spin}$ when $\sigma_x$ and $\sigma_y$ are not allowed at the same site.\footnote{To see this, it is convenient to perform a unitary transformation that maps $\hat{\sigma}_{a,y}$ and $\hat{\sigma}_{a,z}$ to $\hat{\sigma}_{a,z}$ and $-\hat{\sigma}_{a,y}$. 
Then, The Hamiltonian $\hat{H}=\sum J_{abcd}\hat{O}_a\hat{O}_b\hat{O}_c\hat{O}_d$ is mapped to $\hat{H}'=\sum J_{abcd}\hat{O}'_a\hat{O}'_b\hat{O}'_c\hat{O}'_d$, where $\hat{O}'_{2a-1}=\hat{\sigma}_{a,x}$ and $\hat{O}'_{2a}=\hat{\sigma}_{a,z}$ are real and symmetric. We can see that $\hat{H}'$ is real and symmetric if $\hat{O}'_{2a-1}$ and $\hat{O}'_{2a}$ are forbidden to couple directly. The operator $\hat{\Gamma}=\hat{\sigma}_{1,z}\otimes\hat{\sigma}_{2,z}\otimes\cdots\otimes\hat{\sigma}_{N_{\rm spin},z}$ is mapped to $\hat{\Gamma}'=(-1)^{N_{\rm spin}}\hat{\sigma}_{1,y}\otimes\hat{\sigma}_{2,y}\otimes\cdots\otimes\hat{\sigma}_{N_{\rm spin},y}$, which is real and symmetric if $N_{\rm spin}$ is even. 
Therefore, if $N_{\rm spin}$ is even, $\hat{H}'_\pm\equiv\frac{1\pm\hat{\Gamma}'}{2}\hat{H}'$ are real and symmetric, and hence, they are in the GOE universality class. When $N_{\rm spin}$ is odd, there is no specific structure, and hence, we observe the GUE universality class.} 

We could also consider the random $q$-local coupling of $\sigma_{a,x}$, $\sigma_{a,y}$, and $\sigma_{a,z}$ with $a=1,2,\cdots,N_{\rm spin}$. Such a model could be called SpinXYZ$q$. 
The density of states for this model has been studied in Ref.~\cite{Erd_s_2014}.
\subsubsection*{Complex model}
The analog of complex fermions are $\hat{O}_{2a-1}\pm i\hat{O}_{2a}$. By using them, we can define the analog of the complex SYK model.  

\subsubsection*{Coupled SYK-like models}
We can also prepare multiple copies of the SpinXY model and couple them. 
A particularly interesting model of this kind would be the analog of the coupled SYK model~\cite{Maldacena:2018lmt} which could be used to study the traversable wormhole~\cite{Gao:2016bin}.
Note that the traversable wormhole is a promising target of experimental quantum gravity via holography~\cite{Brown:2019hmk}, and there has been an attempt to study the SYK model on a quantum device in this context~\cite{Jafferis:2022crx}. 

\subsubsection*{Qudit models}
We can define a model replacing Pauli operators with spin-$s$ representations with $s>\frac{1}{2}$ and correspondingly replace qubits with qudits as the fundamental quantum registers for quantum simulations.

\section{Density of states}\label{sec:DoS}
\begin{figure}
    \includegraphics[width=8cm]{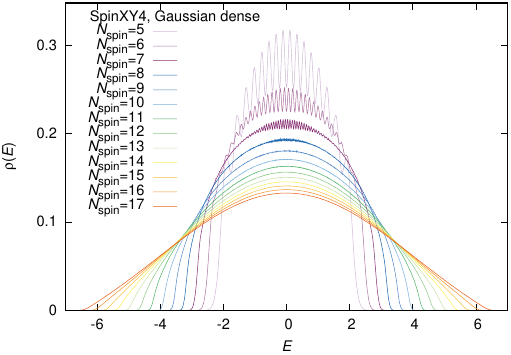}
    \includegraphics[width=8cm]{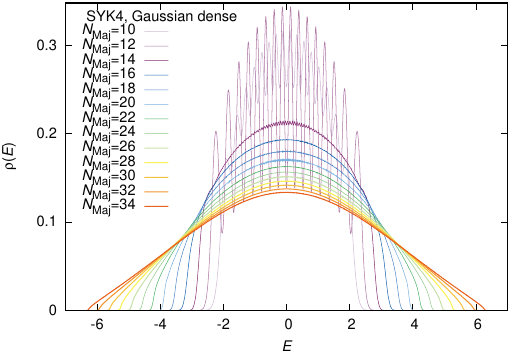}
    \caption{
    The normalized density of states for SpinXY4 (left) and SYK4 (right).
    The contributions of the two parity sectors are not separated.
    The number of samples is $2^{28-N_\mathrm{spin}}$ except for SpinXY4 with $N_\mathrm{spin}=17$ and for SYK4 with $N_\mathrm{Maj}=34$.
    }\label{Fig:DoS-spinXY4-SYK4}
\end{figure}

\begin{figure*}
    \centering
   \scalebox{1.8}{
    \includegraphics{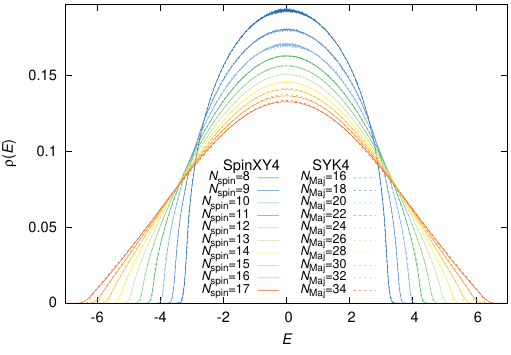}}
    \caption{
    Density of states from $N_{\rm Maj}=2N_{\rm spin}=16$ to $34$. 
    We can see that SpinXY4 and SYK4 have almost the same distribution except for a small discrepancy near the edges. See Fig.~\ref{fig:density-of-state-spin-vs-SYK-edge} for the zoom-in picture near the lower edge. The contributions of the two parity sectors are not separated.}
    \label{fig:density-of-state-spin-vs-SYK}
\end{figure*}

In this section, we define the density of states (DoS) by taking the average over many samples with different random couplings $J_{abcd}$. Practically, we introduce a binning separation of the energy spectrum and count the number of energy levels in each bin. When we combine many samples, we can take a very fine binning width (due to the large statistics of counts). In Fig.~\ref{Fig:DoS-spinXY4-SYK4} we show the DoS obtained in this way. We can see similar shapes across different system sizes. In Fig.~\ref{fig:density-of-state-spin-vs-SYK}, we compared SpinXY4 and SYK4 in the same panel. The two densities are almost indistinguishable, except for a tiny discrepancy near the edges. 

Note that we did not separate the two parity sectors to obtain these results. Whether we separate or not the two parity sectors, we see almost identical densities.

\subsubsection*{Edge of the energy spectrum}\label{sec:edge}
\begin{figure*}
    \centering
    \scalebox{1.8}{
    \includegraphics{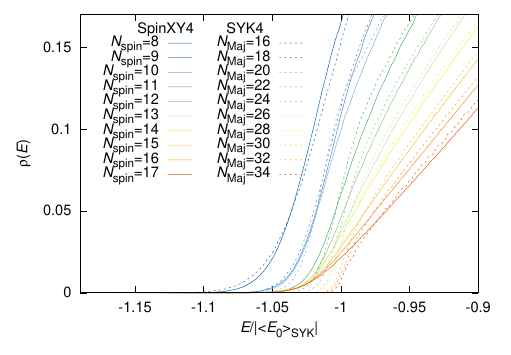}}
    \caption{
    Density of states from $N_{\rm Maj}=2N_{\rm spin}=16$ to $34$ near the edge. The horizontal axis is $E/|\langle E_0\rangle_{\rm SYK}|$. The contributions of the two parity sectors are not separated. 
    }
    \label{fig:density-of-state-spin-vs-SYK-edge}
\end{figure*}

Let us look closely at the edge of the spectrum, where small deviations between the two models are apparent. Fig.~\ref{fig:density-of-state-spin-vs-SYK-edge} is a zoomed-in view of the lower edge of the DoS  from $N_{\rm Maj}=2N_{\rm spin}=16$ to $34$. The horizontal axis is $E/|\langle E_0\rangle_{\rm SYK}|$. As $N_{\rm Maj}$ increases we see a small but clear discrepancy between SpinXY4 and SYK4.  

\begin{figure}
    \centering
    \includegraphics[width=7cm]{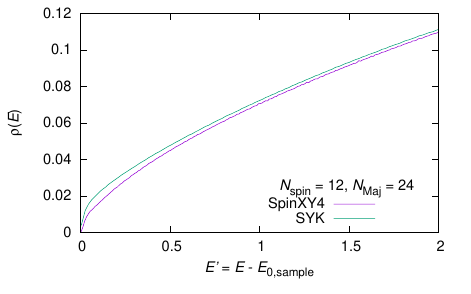}
    \includegraphics[width=7cm]{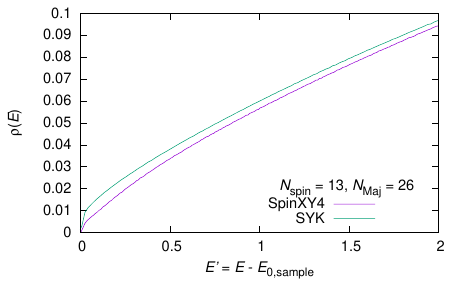}
    \includegraphics[width=7cm]{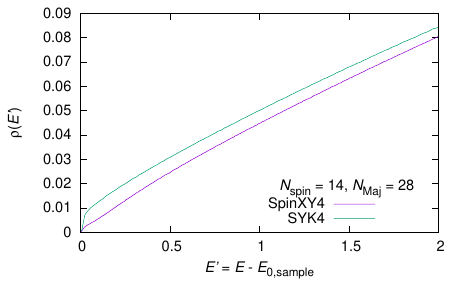}
    \includegraphics[width=7cm]{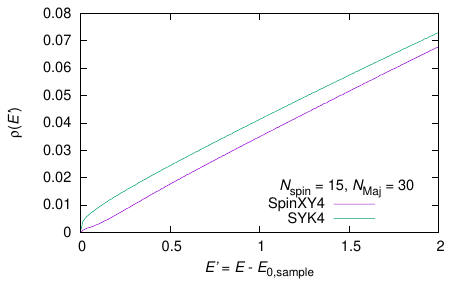}
    \caption{
    The density of $E'=E-E_0$ (sample-by-sample subtraction) for SpinXY4 and SYK4 for $N_{\rm Maj}=2N_{\rm spin}=24, 26, 28, 30$.  
    For both SpinXY4 and SYK4, parity-even sector is used.}
    \label{fig:density-of-state-spin-vs-SYK-edge-1}
\end{figure}

\begin{figure}
    \centering
    \includegraphics[width=7cm]{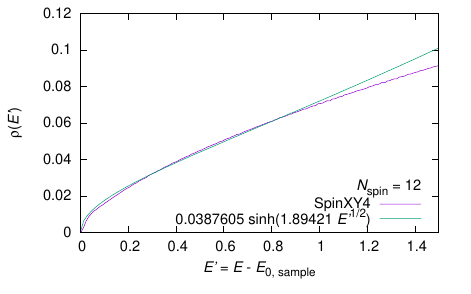}
    \includegraphics[width=7cm]{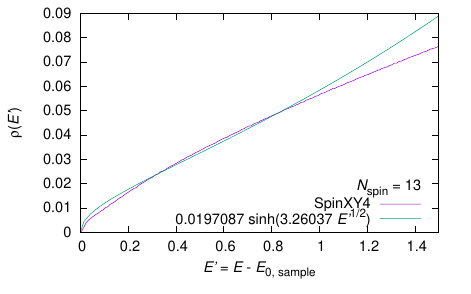}
    \includegraphics[width=7cm]{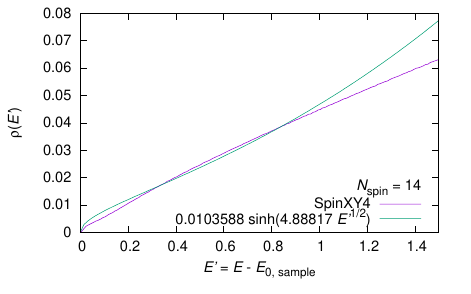}
    \includegraphics[width=7cm]{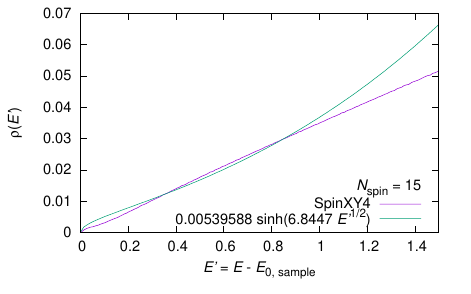}
    \caption{
   The density of $E'=E-E_0$ (sample-by-sample subtraction) for SpinXY4 for $N_{\rm spin}=12, 13, 14, 15$ and fit by $\rho(E')=A\sinh(B\sqrt{E})$ near the edge. The fit is sensitive to the fit range.}
    \label{fig:density-of-state-spin-vs-SYK-edge-2}
\end{figure}

For the SYK model, the DoS behaves as $\rho(E)\simeq A\sinh(B\sqrt{(E-C)})$ near the lower edge, where $C=\langle E_0\rangle$, and $A, B, C$ were estimated analytically~\cite{Cotler:2016fpe,Garcia-Garcia:2016mno,Garcia-Garcia:2017pzl}. A natural question is whether the SpinXY4 shows a similar pattern. A nontrivial technical issue here is that the smallest eigenvalue tends to have a large fluctuation at finite $N$. To deal with this issue, we consider the distribution of $E_i'=E_i-E_0$ ($i=1,2,\cdots$) \cite{Garcia-Garcia:2017pzl} for each parity sector. Note that $E_0$ is subtracted in a sample-by-sample fashion. This option could remove sample-by-sample fluctuations of $E_0$. The distribution of $E'$ is more relevant than that of $E$ when we consider the low-temperature region with the quenched averaging. In Fig.~\ref{fig:density-of-state-spin-vs-SYK-edge-1}, we plotted the density of $E'$ for SpinXY4 and SYK4. While we can see sharp edges for both models, the discrepancy grows as $N$ increases. In Fig.~\ref{fig:density-of-state-spin-vs-SYK-edge-2}, we tried to fit the density of $E'$ for SpinXY4 by $\rho(E')=A\sinh(B\sqrt{E'})$. Although this fit ansatz is not bad for $N_{\rm spin}=12$ and $13$, we do not find a nice fit for $N_{\rm spin}=14$ and $15$. 

\section{Level correlations}\label{sec:Comparison_with_RMT}
In this section, we compare the correlation in the energy spectrum with that of Random Matrix Theory (RMT). 
We will study two sectors corresponding to $\gamma\equiv\prod_{a=1}^{N_{\rm spin}}s_a=\pm 1$ separately. Unlike SYK4, we do not find eigenvalue degeneracy within each sector nor between the two sectors.
We do observe agreement with RMT except for a small number of low-lying eigenvalues. Such an agreement suggests that this model is ergodic rather than in a spin-glass phase. (See e.g., Refs.~\cite{Barney:2023idq,Winer:2022ciz} for the spectral analysis of spin glass.) 
As we will see in Sec.~\ref{sec:SFF}, the spectral form factor of our model resembles that of SYK4. This implies a very strongly chaotic nature of the model.  

\subsection{Nearest-neighbor level spacing}\label{sec:level-spacing}
To compute the eigenvalues, we can utilize the block-diagonal structure of the Hamiltonian, i.e., we can diagonalize $2^{N_{\rm spin}-1}\times 2^{N_{\rm spin}-1}$ blocks corresponding to $\gamma=\pm 1$ separately. 
In each sector, we sorted energy eigenvalues in increasing order as 
$(E_0\equiv)E_1<E_2<\cdots<E_{2^{N_\mathrm{spin}-1}}$. 
The nearest-neighbor level spacing is defined by $s_i \equiv E_{i+1}-E_{i}$. To compare it with RMT, we need to unfold the spectrum. Here we use the fixed-$i$ unfolding~\cite{Gharibyan:2018fax}, i.e., we define the unfolded spacing $\tilde{s}_i$ by $\tilde{s}_i=s_i/\langle s_i\rangle_J$ for each $i$.

\begin{figure}
    \centering
    \scalebox{0.95}{
     \includegraphics{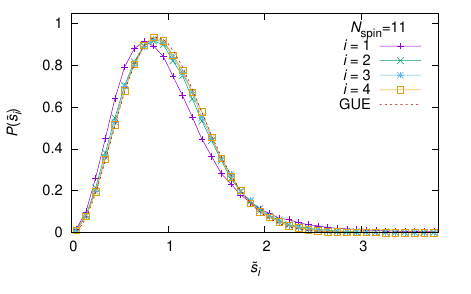}}
    \scalebox{0.95}{
     \includegraphics{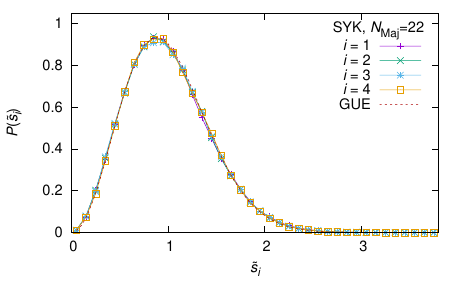}}
    \scalebox{0.95}{
     \includegraphics{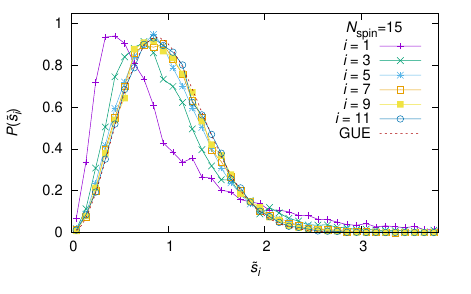}}
    \scalebox{0.95}{
     \includegraphics{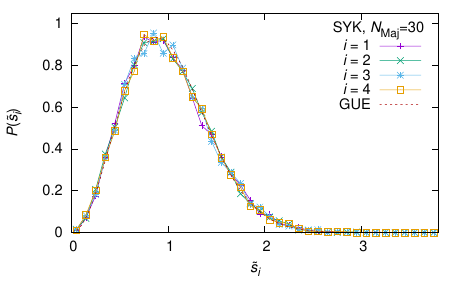}}
    \caption{
          Distribution of the unfolded level spacing $P(\tilde{s}_i)$ for $i=1,2,3,\cdots$. 
          SpinXY4, $N_{\rm spin}=11$ (top, left), SYK4, $N_{\rm Maj}=22$ (top, right), SpinXY4, $N_{\rm spin}=15$ (bottom, left), SYK4, $N_{\rm Maj}=30$ (bottom, right). 
          Only the parity-even sector was used. 
          In SpinXY4, the eigenvalues in the parity-even and odd sectors are not correlated but the plots for the two sectors are indistinguishable. In SYK4, parity-even and parity-odd sectors have the same eigenvalues when $N_{\rm spin}=N_{\rm Maj}/2$ is odd. 
    }
    \label{fig:gap-each-i}
\end{figure}

In Fig.~\ref{fig:gap-each-i}, the distribution of the unfolded level spacing $P(\tilde{s}_i)$ is plotted for several values of $i$. For $N_{\rm spin}$. For $N_{\rm spin}=11$, although a significant difference from RMT can be seen only for $i=1$, we see almost no difference from RMT at $i\ge 2$. For $N_{\rm spin}=15$, we see a larger deviation from RMT at small $i$. However, the agreement with RMT is not bad already at $i=4$ and it is hard to see a difference from RMT at $i\ge 10$. For the SYK4, a good agreement with RMT is observed even for $i=1$~\cite{Gur-Ari:2018okm}.

\subsection{Neighboring gap ratio}\label{sec:gap-ratio}
\begin{figure}
    \centering
   {
    \includegraphics{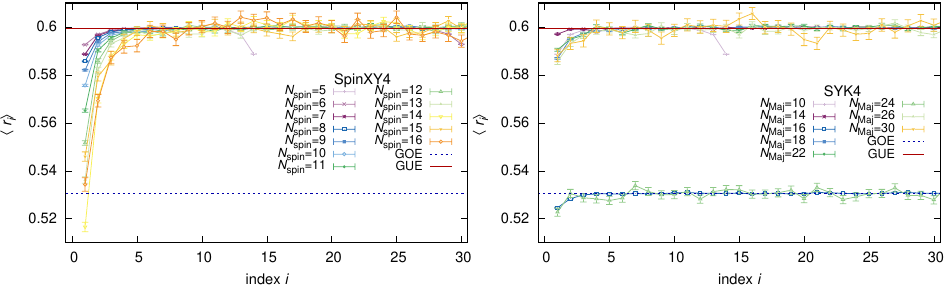}}
    \caption{Neighboring gap ratio $\langle r_i\rangle$ for $i=1,2,3,\cdots,30$, SpinXY4 (left) and SYK4 (right).
    For both, only the parity-even sector was used. For SYK4, $N_\mathrm{Maj}=12,20,28$ that correspond to Gaussian symplectic ensemble (GSE) are omitted.
       }
    \label{fig:r-each-i}
\end{figure}

By using the unfolded level spacing $\tilde{s}_i$, we define the neighboring gap ratio $r_i$ as
\begin{align}
    r_i=\frac{
    \min(\tilde{s}_i,\tilde{s}_{i+1})
    }{
    \max(\tilde{s}_i,\tilde{s}_{i+1})
    }\, .
\end{align}
In the left panel of Fig.~\ref{fig:r-each-i}, we plotted $\langle r_i\rangle$ for SpinXY4, $i=1,2,3,\cdots,30$, $N_{\rm spin}=11,\cdots,16$. Good agreement with RMT (the GUE universality class)~\cite{Atas2013distribution} is observed at $i\ge 4$. 
In the right panel, the same quantities for SYK4 are plotted for the values of $N$ corresponding to GUE. Again, a good agreement with RMT is observed at $i\ge 4$. 

Note that the gap ratio can be sensitive to the unfolding near the edges of the energy spectrum. By using the unfolded level spacings, we can see good agreement even near the edge of the spectrum.

\subsection{Spectral form factor}\label{sec:SFF}
\begin{figure}
    \centering
    \includegraphics[width=8cm]{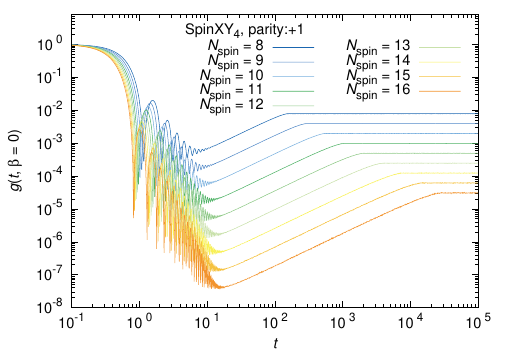}
    \includegraphics[width=8cm]{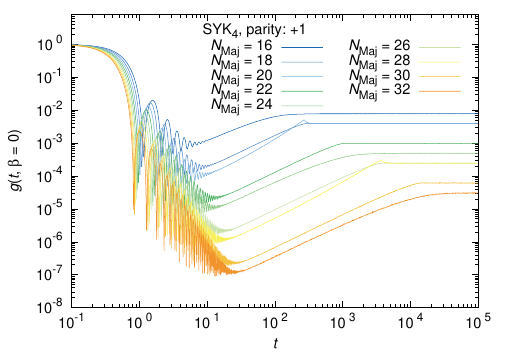}
    \caption{
The spectral form factor for SpinXY4, $N_\mathrm{spin} = 8,9,\ldots,16$ (left) and that for SYK4, $N_{\rm Maj}=2N_\mathrm{spin}=16,18,\ldots,32$ (right). 
Only the parity-even sector is used. Note that SYK4 has a two-fold degeneracy in eigenvalues in each parity sector when $N_{\rm Maj}\equiv 4$ mod 8 and such a degeneracy shifts the height of the plateau by factor 2. 
The number of samples is $2^{28-(N_\mathrm{spin})}$ for both SpinXY4 and SYK4.
     }
    \label{fig:XY4-GtZb0-Ns5-13}
\end{figure}

A convenient quantity to see the correlation of energy eigenvalues in a wider energy band is the spectral form factor (SFF). 
The SFF can be defined for each parity sector as
\begin{align}
    g_{\gamma=\pm 1}(t,\beta)=\frac{\langle\vert Z_{\gamma=\pm 1}(t,\beta)\vert^2\rangle_J}{\langle\vert Z_{\gamma=\pm 1}(0,\beta)\vert^2\rangle_J}\, , 
\end{align}
where
\begin{align}
    Z_{\gamma=\pm 1}(t,\beta)\equiv Z_{\gamma=\pm 1}(\beta+it) = \sum_j \exp\left(-(\beta+it)E_j\right)\, . 
\end{align}
Here the sum over states $j$ is taken in $\gamma=+1$ or $\gamma=-1$ sector. 
The SFF starts with $1$ at $t=0$ and shows the slope, dip, ramp, and plateau. The ramp and plateau are universal among chaotic systems. If the ramp is longer (equivalently, if the onset of the ramp is earlier), the energy spectrum agrees with RMT in the wider energy band. 
We plotted $g(t,\beta=0)$ for our model in Fig.~\ref{fig:XY4-GtZb0-Ns5-13}. We can compare it with the same quantity for the SYK model. We can see similar long ramps.  

\begin{figure}
    \centering
    \includegraphics[width=8cm]{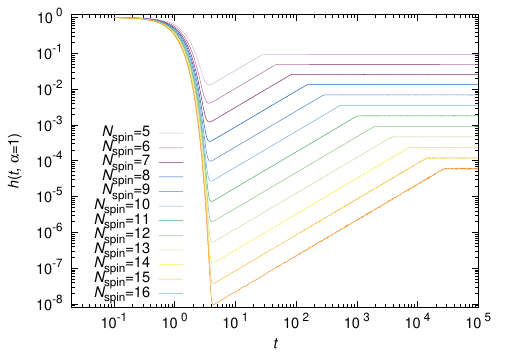}
    \includegraphics[width=8cm]{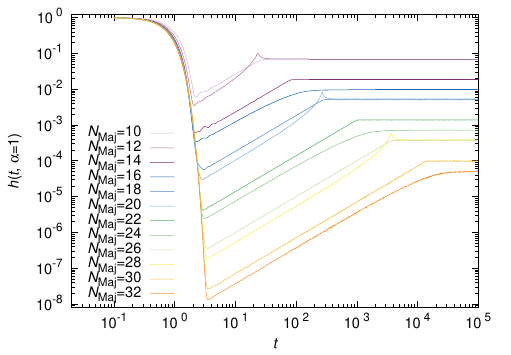}
    \caption{
    The modified spectral form factor for SpinXY4, $N_\mathrm{spin} = 5,6,\ldots,16$ (left) and SYK4, $N_{\rm Maj}=2N_\mathrm{spin}=10,12,\ldots,32$ (right).
    Only the parity-even sector is used.
    The number of samples is $2^{28-(N_\mathrm{spin})}$ for both SpinXY4 and SYK4.
     }
    \label{fig:XY4-HtZa1-Ns5-13}
\end{figure}

The onset of the ramp can be hidden by the slope. To see the onset of the ramp more accurately, a modified spectral form factor $h(\alpha,t,\beta)$ defined by~\cite{Stanford-unpublished,Gharibyan:2018jrp}
\begin{align}
    h_{\gamma=\pm 1}(\alpha,t,\beta)=\frac{\langle\vert Y_{\gamma=\pm 1}(\alpha,t,\beta)\vert^2\rangle_J}{\langle\vert Y_{\gamma=\pm 1}(\alpha,0,\beta)\vert^2\rangle_J}\, ,  
\end{align}
where
\begin{align}
    Y_{\gamma=\pm 1}(\alpha,t,\beta)\equiv \sum_j \exp\left(-\alpha E_j^2-(\beta+it)E_j\right)\, , 
\end{align}
is useful. By tuning a parameter $\alpha$ appropriately, the slope can fall much more quickly and the hidden part of the ramp can be revealed. $h_{\gamma=\pm 1}(\alpha=1,t,\beta=0)$ is plotted in Fig.~\ref{fig:XY4-HtZa1-Ns5-13}. 

\subsubsection{Spectral Form Factors with and without separating parity sectors}
In the above, we defined the SFF for each parity sector. We could also combine two sectors. Specifically, by using $Z_{\rm full}=Z_{\gamma=+1}+Z_{\gamma=-1}$, we could define $g_{\rm full}(t,\beta)\equiv \frac{\langle|Z_{\rm full}(t,\beta)|^2\rangle_J}{\langle|Z_{\rm full}(0,\beta)|^2\rangle_J}$ as the `full' SFF.

\begin{figure}
    \centering
    \includegraphics{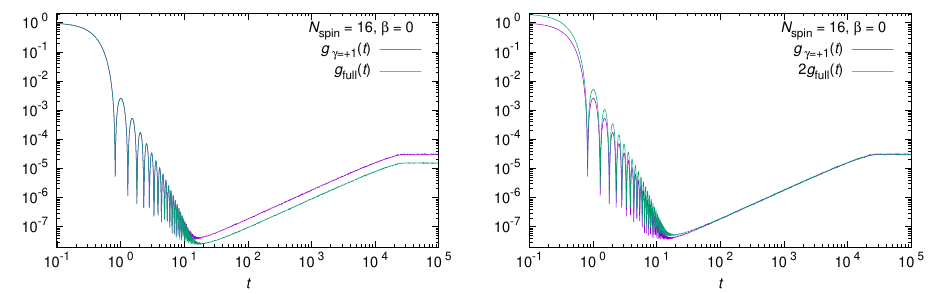}
    \caption{
The spectral form factor for SpinXY4 for $\beta=0$, $N_\mathrm{spin}=16$ (4096 samples).
[Left] $g_{\rm full}(t,\beta)=g_{\gamma=\pm 1}(t,\beta)$ can be seen at an early time. [Right] $g_{\rm full}(t,\beta)=\frac{1}{2}g_{\gamma=\pm 1}(t,\beta)$ can be seen at late time.}
    \label{fig:SFF-full-vs-each-gamma}
\end{figure}

In Fig.~\ref{fig:SFF-full-vs-each-gamma}, we plotted $g_{\rm full}(t,\beta)$ and $g_{\gamma=\pm 1}(t,\beta)$. We can see that
\begin{align}
g_{\rm full}(t,\beta)
=
g_{\gamma=\pm 1}(t,\beta)
\label{full-SFF-early-time}
\end{align}
at an early time and
\begin{align}
g_{\rm full}(t,\beta)
=
\frac{1}{2}\cdot g_{\gamma=\pm 1}(t,\beta)
\label{full-SFF-late-time}
\end{align}
at late time. It can be understood as follows. 
\begin{itemize}
\item 
At early time, $Z_{\gamma=\pm 1}$ are self-averaging and take the same value. Therefore, $Z_{\rm full}=2Z_{\gamma=+1}=2Z_{\gamma=-1}$ up to $1/N_{\rm spin}$-corrections without the average over random couplings, which implies \eqref{full-SFF-early-time}. 

\item 
To see the late-time behavior, we use 
$|Z_{\rm full}|^2=|Z_{\gamma=+1}|^2+|Z_{\gamma=-1}|^2+2{\rm Re}(Z_{\gamma=+1}Z_{\gamma=-1}^\ast)$. The first two terms on the right-hand side are described by RMT with matrix size $2^{N_{\rm spin}-1}\times 2^{N_{\rm spin}-1}$. The third term vanishes because there is no correlation between two parity sectors and hence $Z_{\gamma=\pm 1}$ fluctuates around zero independently, averaging to zero: $\langle Z_{\gamma=+1}Z_{\gamma=-1}^\ast\rangle_J=0$. 
Therefore, the late-time behavior of $\langle |Z_{\rm full}|\rangle_J$ coincides with that of RMT. The factor-2 difference in \eqref{full-SFF-late-time} is explained by the difference in the dimension of the full and parity-fixed Hilbert spaces. 

\end{itemize}

\subsubsection{Comparison with SYK at all time scales}\label{sec:SFF-all-time scale}
As we have seen above, the late-time features of the SFF capture the fine-grained energy-level correlations. On the other hand, at the early time, the SFF is sensitive to the density of states. Therefore, the observation so far indicates that the SFF of SpinXY4 resembles that of SYK4 closely both at an early time and at a late time. Now we would like to ask if the similarity can be observed at all time scales. 

For quantitative agreement at the late time, we choose $N$ such that $N_{\rm spin}=N_{\rm Maj}/2$ is odd because then both SpinXY4 and SYK4 are in the GUE universality class and hence we can expect the precise agreement at a late time. With such a choice of $N$, there is a two-fold degeneracy in the energy eigenvalues in SYK4. Therefore, there are $2^{N_{\rm spin}-1}$ independent eigenvalues used in the SFF. If we keep only one of the parity sectors in the spin model, the numbers of eigenvalues match. Therefore, we compare $g_{\gamma=\pm 1}(t,\beta)$ in SpinXY4 and $g(t,\beta)$ in SYK4.

\begin{figure}
    \centering
    \includegraphics[width=8cm]{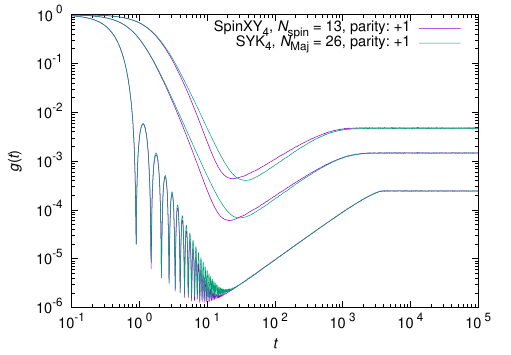}
    \includegraphics[width=8cm]{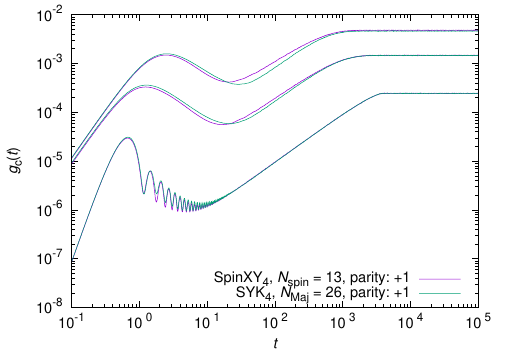}
    \includegraphics[width=8cm]{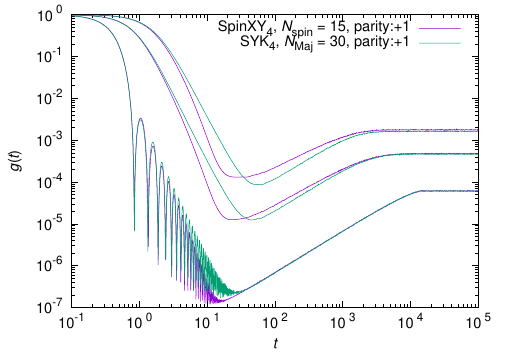}
    \includegraphics[width=8cm]{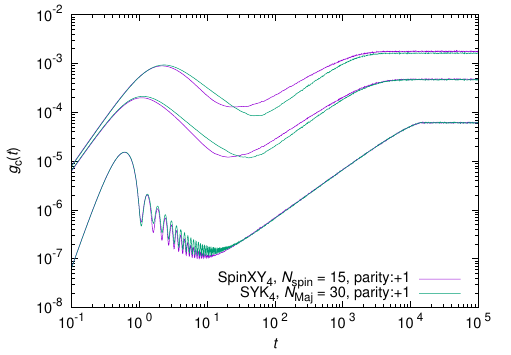}    
    \caption{
    $g(t)$ (left) and $g_\mathrm{c}(t)$ (right) for SpinXY4 and SYK4 are compared.
    The results for $\beta=0,1,2$ are plotted from bottom to top. 
    [Top] $N_\mathrm{spin}=13$. $32768$ samples are used for both models.
    [Bottom] $N_\mathrm{spin}=15$. $8192$ samples are used for both models. 
    The parity-even sector is used for both SpinXY4 and SYK4. 
     }
    \label{fig:SFF-SYK-vs-spin-N26}
\end{figure}

In Fig.~\ref{fig:SFF-SYK-vs-spin-N26}, we plot the spectral form factor for $N_{\rm Maj}=2N_{\rm spin}=26$ and $30$, $\beta=0,1$, and $2$. In addition to $g(t,\beta)$, we plot the `connected part' defined by 
\begin{align}
g_{\rm c}(t,\beta)\equiv
\frac{\langle|Z(t,\beta)|^2\rangle_J-|\langle Z(t,\beta)\rangle_J|^2}{\langle|Z(0,\beta)|^2\rangle_J}\, . 
\end{align}
The agreement is strikingly good, although a small discrepancy is visible around the dip. 

\section{Edwards-Anderson parameter}\label{sec:EA}

The Edwards-Anderson parameter \cite{S_F_Edwards_1975} is a standard tool to see if a given system has a spin-glass phase or not. In this section, we study the generalized Edwards-Anderson parameter $q_\mathrm{gEA}(j)$, here defined for the $j$-th lowest energy normalized eigenstates $\ket{E_j}^{(\mathrm{E}) , (\mathrm{O})}$ as~\cite{swingle2023bosonic}
\begin{equation}
q_\mathrm{gEA}(j)
= \frac{1}{N_\mathrm{spin}}
    \sum_{i}
    \sum_{\alpha=x,y}
    \left\vert
    \bra{\psi_j^{(\mathrm{O})}}
    \hat{\sigma}_{i,\alpha}
    \ket{\psi_j^{(\mathrm{E})}}
    \right\vert^2\, .
\end{equation}
Note that we do not include $\alpha=z$ in the sum because $\bra{\psi_j^{(\mathrm{O})}}
    \hat{\sigma}_{i,z}
    \ket{\psi_j^{(\mathrm{E})}}=0$ due to the parity conservation.
We also study $q_\mathrm{zEA}$ defined by 
\begin{equation}
q_\mathrm{zEA}(j)
= \frac{1}{N_\mathrm{spin}}
    \sum_{i}
    \left\vert
    \bra{\psi_j}
    \hat{\sigma}_{i,z}
    \ket{\psi_j}
    \right\vert^2\, .
\end{equation}
Numerically, we observed that $q_\mathrm{zEA}$ takes a nonzero value only for odd $N_\mathrm{spin}$.

\begin{figure}
    \includegraphics[width=0.5\linewidth]{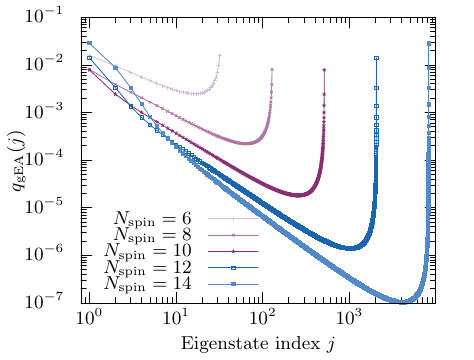}
    \includegraphics[width=0.5\linewidth]{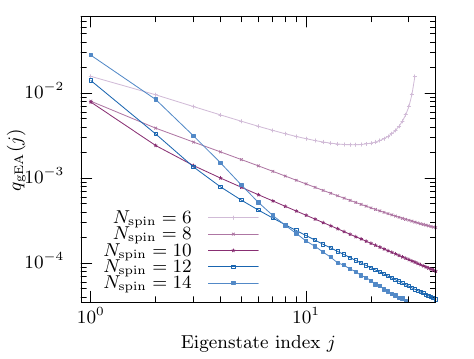}
    \includegraphics[width=0.5\linewidth]{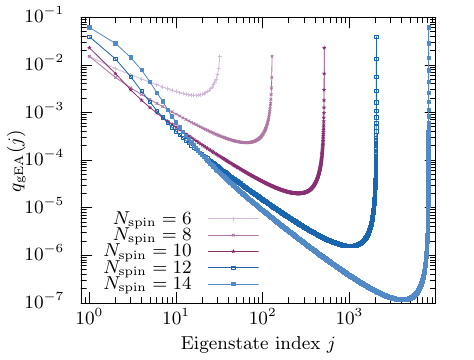}
    \includegraphics[width=0.5\linewidth]{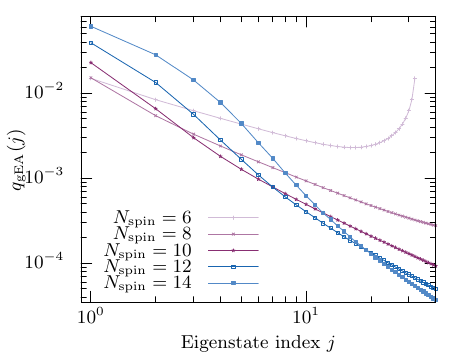}
    \caption{
    The averaged values of Edwards-Anderson parameter plotted as function of the eigenstate index for $N_\mathrm{spin}=6,8,10,12,14$. $2^{28-N_\mathrm{spin}}$ samples are used. The eigenstate index, $j=1,2,\ldots, 2^{N_\mathrm{spin}-1}$ is allocated in the increasing order of the eigenstate energy for each parity sector, and data from both parity sectors are used.
    Top: terms with $\eta_{abcd}>0$ are allowed in \eqref{spin_Hamiltonian}.
    Bottom: $J_{abcd}$ is set to zero if $\eta_{abcd}>0$.
    }
    \label{fig:genEA}
\end{figure}

In Fig.~\ref{fig:genEA} we plot the value of $q_\mathrm{gEA}(j)$ as a function of the eigenstate index $j\in[1,2^{N_\mathrm{spin}-1}]$.
For clarity we only plot the results for even $N_\mathrm{spin}$, however the results for odd $N_\mathrm{spin}$ qualitatively agree with the even $N_\mathrm{spin}$ case as we see below. Due to the symmetry concerning the overall sign of the Hamiltonian, the distributions of $q_\mathrm{gEA}(j)$ and $q_\mathrm{gEA}(2^{N_\mathrm{spin}-1}+1-j)$ are identical.
At $N_\mathrm{spin}\le 14$ we observed the following pattern:
\begin{itemize}
    \item $q_\mathrm{gEA}(j)$ at small and fixed $j$ increases as a function of $N_\mathrm{spin}$, indicating that the lowest energy eigenstates behave as spin glass states.
    \item For $j>\mathcal{O}(10^1)$, $q_\mathrm{gEA}(j)$ decreases as a function of $N_\mathrm{spin}$. However, it is possible that $q_{\rm gEA}$ eventually increase with $N_\mathrm{spin}$ at any fixed $j$, if $N_\mathrm{spin}$ becomes sufficiently large. 
    \item At fixed $N_\mathrm{spin}$, $q_\mathrm{gEA}(j)$ shows a power-law decay as a function of the eigenstate index, until $j$ reaches $\approx 2^{N_\mathrm{spin}-2}$.
    \item The smallest value of $q_\mathrm{gEA}$ decreases exponentially as $N_\mathrm{spin}$ is increased.
\end{itemize}
    Removal of terms with $\eta_{abcd}>0$, where multiple operators acting on the same spin are chosen, affects the value of $q_\mathrm{gEA}(j)$ slightly but does not qualitatively change the pattern above.

\begin{figure}
    \includegraphics[width=0.5\linewidth]{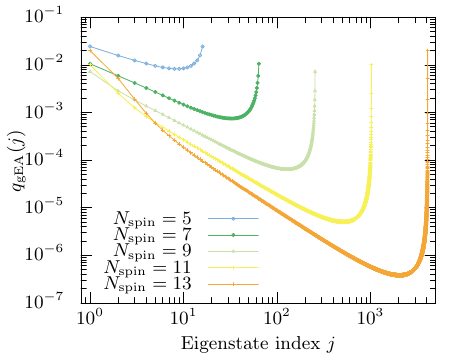}
    \includegraphics[width=0.5\linewidth]{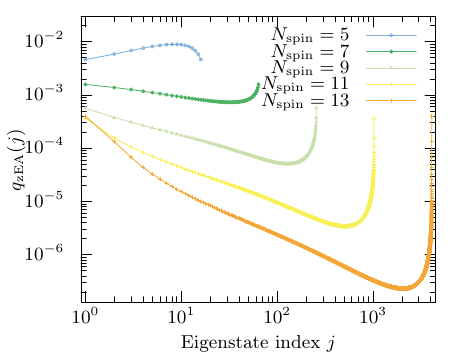}
    \includegraphics[width=0.5\linewidth]{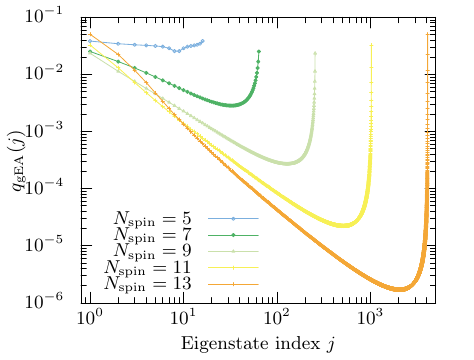}
    \includegraphics[width=0.5\linewidth]{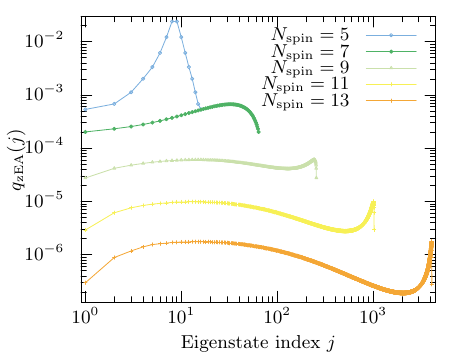}
    \caption{
    The averaged values of (left) the generalized Edwards-Anderson parameter $q_\mathrm{gEA}$ and (right) the Edwards-Anderson parameter $q_\mathrm{zEA}$ plotted as functions of the eigenstate index for $N_\mathrm{spin}=5,7,9,11,13$. $2^{28-N_\mathrm{spin}}$ samples are used. The eigenstate index, $j=1,2,\ldots, 2^{N_\mathrm{spin}-1}$ is allocated in the increasing order of the eigenstate energy for each parity sector, and data from both parity sectors are used.
    Top: terms with $\eta_{abcd}>0$ are allowed in \eqref{spin_Hamiltonian}.
    Bottom: $J_{abcd}$ is set to zero if $\eta_{abcd}>0$.
    }
    \label{fig:zEA}
\end{figure}

In Fig.~\ref{fig:zEA}, we plot $q_\mathrm{gEA}$ and $q_\mathrm{zEA}$ for odd $N_\mathrm{spin}$. We observe that $q_\mathrm{gEA}$ behaves similarly to the case of even values of $N_\mathrm{spin}$. While $q_\mathrm{zEA}(j)$ shows a similar behavior with a smaller number of $j$ showing increase with $N_\mathrm{spin}$ between $N_\mathrm{spin}=11$ and $N_\mathrm{spin}=13$, it decreases exponentially as $N_\mathrm{spin}$ is increased for all $j$ when terms with $\eta_{abcd}>0$ are removed from the model \eqref{spin_Hamiltonian}.

In summary, $q_\mathrm{gEA}(j)$ suggests some low-energy states are in the spin-glass phase, although $q_\mathrm{zEA}(j)$ suggests the opposite may be the case if terms with $\eta_{abcd}>0$ are suppressed. More studies will be needed to have a conclusive statement. In this context, we note that Ref.~\cite{swingle2023bosonic} studied the Edwards-Anderson parameter for the ground state of the SpinXYZ$q$ model.  They observed a slow decline of the Edwards-Anderson parameter that is consistent with the absence of the glassiness, although the signal could mean that $q=4$ is sitting at the border between spin-glass and ergodic cases in the sense that $q=3$ and $q=5$ respectively exhibit clear growth and decline of the Edwards-Anderson parameter.
Also, see Ref.~\cite{ish2020SYK} that analyzed an analogue of the EA parameter to study the phase structure of the SYK model.

\section{Two-point function}\label{sec:2-pt-fnc}
We consider the two-point function
\begin{align}
    \frac{1}{Z(\beta)}\sum_{E}
    e^{-\beta E}\bra{E}\hat{O}_a(t)\hat{O}_a(0)\ket{E}
    =
    \frac{1}{Z(\beta)}\sum_{E,E'}
    e^{-\beta E+i(E-E')t}
    \left|\bra{E}\hat{O}_a\ket{E'}\right|^2\, . 
\end{align}
Note that we take the sum over all the energy eigenstates from both parity $\gamma=\pm 1$ sectors. We will take the average over random couplings separately for the numerator and denominator. Furthermore, we take the average over $a=1,\cdots,N_{\rm Maj}$. 
Here we consider the annealed average:
\begin{align}
G_{x,y}(t)
\equiv
\frac{1}{2N_{\rm spin}}\cdot
    \frac{1}{\langle Z(\beta)\rangle_J}
    \left\langle
    \sum_{E,E'}
    e^{-\beta E+i(E-E')t}
    \sum_{a=1}^{2N_{\rm spin}}
    \left|\bra{E}\hat{O}_a\ket{E'}\right|^2
    \right\rangle_J\, . 
    \label{def:G_xy}
\end{align}
We will also consider 
\begin{align}
G_{z}(t)
\equiv
\frac{1}{N_{\rm spin}}\cdot
    \frac{1}{\langle Z(\beta)\rangle_J}
    \left\langle
    \sum_{E,E'}
    e^{-\beta E+i(E-E')t}
    \sum_{j=1}^{N_{\rm spin}}
    \left|\bra{E}\hat{\sigma}_{j,z}\ket{E'}\right|^2
    \right\rangle_J\, . 
     \label{def:G_z}
\end{align}

\begin{figure}
    \begin{center}
        \includegraphics{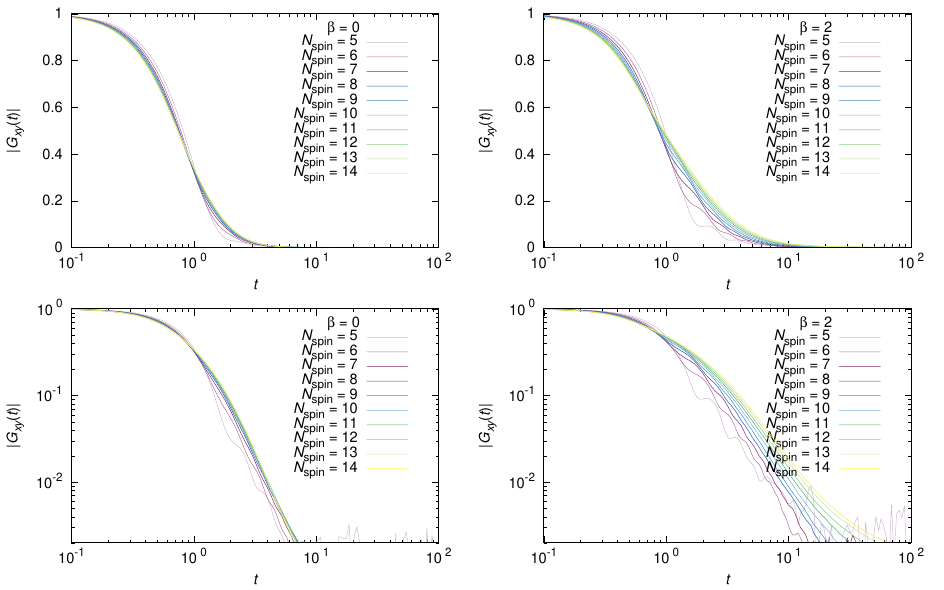}
    \end{center}
    \caption{$|G_{xy}(t)|$ plotted for the SpinXY4 model at $\beta=0$ (left) and $\beta=2$ (right) from top to bottom. The vertical axis is linear (logarithmic) in the upper (lower) plots. 1024 samples are used, and the average over all operators and samples is taken before the absolute value is computed.}
    \label{fig:SpinXY4-2ptXY}
\end{figure}

\begin{figure}
    \begin{center}
        \includegraphics[width=16cm]{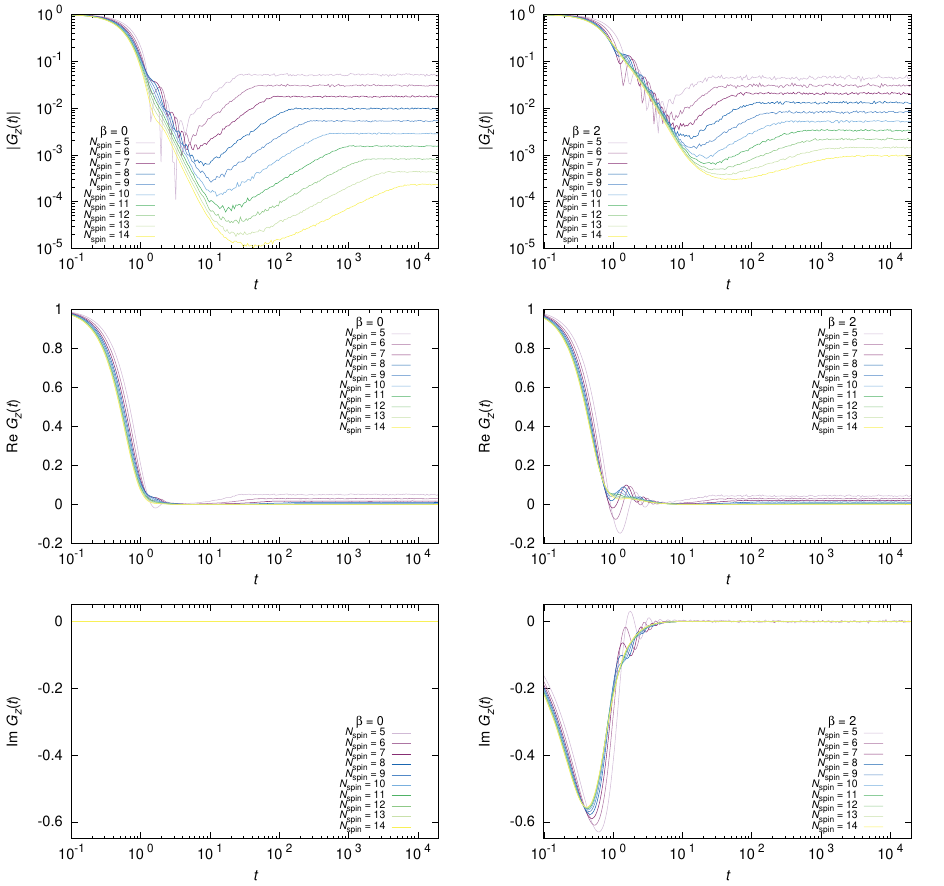}
    \end{center}
    \caption{$|G_z(t)|$, $\mathrm{Re}~G_z(t)$, and $\mathrm{Im}~G_{z}(t)$ for $\beta=0,2$ plotted for the SpinXY4 model. Note that $\mathrm{Im}~G_{z}(t)=0$ holds for $\beta=0$, which is numerically confirmed. 1024 samples are used, and the average over all operators and samples is taken before the absolute value is computed.}
    \label{fig:SpinXY4-2ptZ}
\end{figure}

\begin{figure}
    \begin{center}
        \includegraphics[width=16cm]{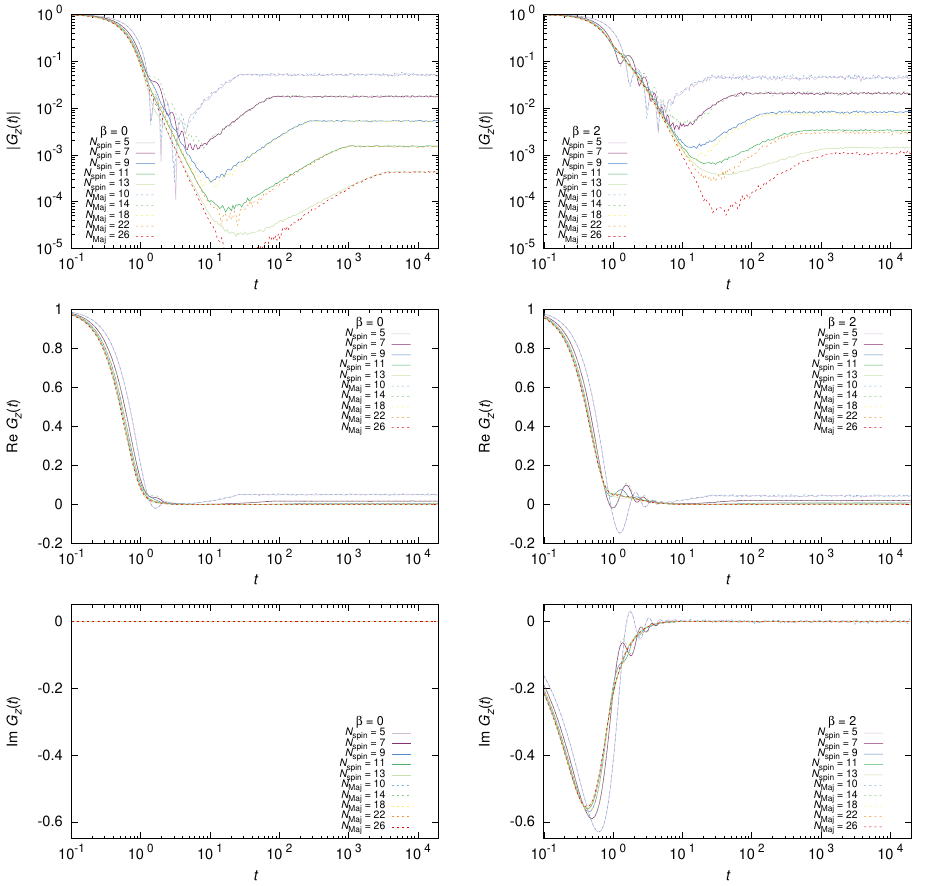}
    \end{center}
    \caption{$|G_z(t)|$, $\mathrm{Re}~G_z(t)$, and $\mathrm{Im}~G_{z}(t)$ for $\beta=0,2$ plotted for the SpinXY4 model ($N_\mathrm{spin}=5,7,9,11,13$) and for the SYK4 model ($N_\mathrm{Maj}=10,14,18,22,26$). Note that $\mathrm{Im}~G_{z}(t)=0$ holds for $\beta=0$, which is numerically confirmed. 1024 samples are used, and the average over all operators and samples is taken before the absolute value is computed.}
    \label{fig:SpinXY4SYK4-2ptZ}
\end{figure}

If the system is chaotic, late-time behaviors of such correlation functions should be understood based on RMT. We can repeat the argument for the SYK model~\cite{Cotler:2016fpe} without a substantial change. 

As for $G_{x,y}(t)$, the operators $\hat{O}_a$ connect states with different parity, 
and hence, $\left|\bra{E}\hat{O}_a\ket{E'}\right|^2$ is nonzero when $\ket{E}$ and $\ket{E'}$ are in different parity sectors. Other than that, it can be approximated by a smooth function of $E-E'$, as suggested by the eigenstate thermalization hypothesis (ETH). As far as the late-time behaviors are concerned, we can approximate it with a constant. 
Then, contributions from two sectors with different parity, which are not correlated, cancel out and we do not see the ramp and plateau. 
As we can see in Fig.~\ref{fig:SpinXY4-2ptXY}, this is indeed the case. We can see a close similarity with two-point function of $\psi_i(t)$ and $\psi_i(0)$ in SYK4 with $N_{\rm Maj}=2N_{\rm spin}=16, 20$ and $24$~\cite{Cotler:2016fpe}. 

As for $G_{z}(t)$, the operators $\hat{\sigma}_{j,z}$ do not change the parity. Therefore, the late-time behavior resembles the sum of SFF in two parity sectors, and hence, we expect the ramp and plateau. In Fig.~\ref{fig:SpinXY4-2ptZ}, we do see such a pattern.

\subsection{Comparison with SYK model at all time scales}
In Sec.~\ref{sec:SFF-all-time scale}, we observed that the spectral form factors from SYK4 and SpinXY4 can be close at all time scales. Let us see if a similar coincidence can be seen for the two-point functions. 

We take $N_{\rm spin}$ odd so that both models are in the GUE universality class. The eigenvalues in two parity sectors in SpinXY4 are not correlated while the eigenvalues in the two parity sectors of the SYK model are paired. Therefore, we compare fixed-parity sectors in SpinXY4 and SYK4, and we choose the operators that do not mix different parity sectors. Specifically, we study the two-point function of $\hat{\sigma}_{a,z}=-i\hat{O}_{2a-1}\hat{O}_{2a}=-i\hat{\chi}_{2a-1}\hat{\chi}_{2a}$, which is $G_z(t)$ defined by \eqref{def:G_z}. 

The results are shown in Fig.~\ref{fig:SpinXY4SYK4-2ptZ}, for $\beta=0,2$. Overall, we find them remarkably similar to each other. For $\beta=0$, we observe good agreement at early and late times, although some discrepancy is visible in between. For $\beta=2$, we can see a small difference at late time as well.

\section{Path-integral approach}\label{sec:path-integral}
We discuss how large-$N$ spin systems can be studied systematically with path integral methods. To develop a systematic large-$N$ (and $1/N$) expansion, the following features are needed:
\begin{enumerate}[label=\roman*)]
    \item \label{item:singlets} An invariant (collective) set of variables $\Phi(a)$ needs to be identified, generally as singlets under a U($N$), O($N$), Sp($2 N$) or S$_N$ group operating on the system.
    \item \label{item:SD} A closed set of Schwinger-Dyson (SD) equations needs to be deduced and/or
    \item \label{item:collective}a collective action describing the $1/N$ dynamics of collective variables $\Phi(a)$ needs to be established.
\end{enumerate}
We note that \cref{item:SD,item:collective} should have an equivalent description in terms of Feynman diagrams, e.g., planar diagrams in matrix models, and bubble diagrams in vector models. For theories of spin degrees of freedom, however, none of the required features \cref{item:singlets,item:SD,item:collective} were obvious so far. We will see shortly that for spin systems the relevant symmetry group is S$_N$, which induces an infinite set of collective variables. This S$_N$ symmetry is featured by expanding the kinetic term in the Lagrangian, thus our formalism described below applies to all spin systems.

Let $S_i^{a} = \frac{1}{2} \sigma_{i,a}$ denote the spin operator, where $a = x, y, z$, and $i = 1, \dots, N_{\rm spin}$ denotes the site index. From here on, we use the letter $N$ instead of $N_{\rm spin}$, i.e., $N=N_{\rm spin}$. Furthermore, we take the variance of the random coupling to be $J^2$. (Previously, we took $J=1$.)
Up to the $1/N$-suppressed terms, the real-time path integral after the disorder average is given by
\begin{align}
    Z_{J} = \int \prod_{i=1}^{N} & \mathcal{D} \vec{S}_i \, \delta\bigl(\vec{S}_i^2 - s^2\bigr) \nonumber \\
    \label{eq:spin_XY4_Z_J}
    \times & \exp\left[
        \ii \int L_{\kinetic} dt - \frac{J^2N}{4} \int dt_1 dt_2 \Biggl(
            \frac{1}{N}\sum_{i=1}^{N} 
            \left(S_i^+(t_1) S_i^-(t_2) + S_i^-(t_1) S_i^+(t_2)\right)
        \Biggr)^4
    \right] \, ,
\end{align}
where $s=\frac{1}{2}$ and $S_i^{\pm} \equiv S_i^x \pm \ii S_i^y$. In the above formula we only consider the $\eta = 0$ sector in the Hamiltonian \eqref{spin_Hamiltonian} since in the large $N$ limit the $\eta > 0$ sector is of lower order in $1/N$ and hence can be dropped. To be more specific, one can see from above that the $\eta = 0$ sector gives a potential of order $\mathcal{O}(N)$. On the other hand, one can show that the $\eta = 1$ and $\eta = 2$ sectors are of order $\mathcal{O}(1)$ and $\mathcal{O}(1/N)$ respectively, and thus are suppressed in the large $N$ limit. The kinetic term in the Lagrangian is
\begin{equation}
    L_{\kinetic} = \frac{\ii}{2} \sum_{i=1}^{N} \frac{S_i^- \dot{S}_i^+ - S_i^+ \dot{S}_i^-}{s + S_i^z}\, .
\end{equation}
This term does not have the O($N$) symmetry as opposed to the SYK model \cite{Jevicki:2016bwu}, while the S$_N$ symmetry is manifest. We can use the constraint to write $S^z = \sqrt{s^2 - S^+ S^-}$ and expand the denominator into a Taylor series as 
\begin{equation}
    L_{\kinetic} = \frac{\ii}{2 s}\left[
    \sum_{i=1}^{N} (S_i^- \dot{S}_i^+ - S_i^+ \dot{S}_i^-) \left(
        \frac{1}{2} + \frac{1}{8 s^2} S_i^+ S_i^- + \frac{1}{16 s^4} S_i^+ S_i^- S_i^+ S_i^- 
        + \dots
    \right) \right]\, . 
\end{equation}
This expression motivates us to use the S$_N$-singlet multi-time collective variables 
\begin{equation}
    \label{eq:S_N_invariants}
    \Phi_{L}(t_1, t_2, \dots, t_L; t^\prime_1, t^\prime_2, \dots, t^\prime_L) = 
        \frac{1}{N}\sum_{i=1}^N S_i^+(t_1) S_i^+(t_2) \cdots S_i^+(t_L) S_i^-(t^\prime_1) S_i^-(t^\prime_2) \cdots S_i^-(t^\prime_L) \, ,
\end{equation}
with $L$ being the length of the sequence (number of pairs of $S_i^+ S_i^-$). The multi-time labels of these collective variables are themselves identical under the S$_L$-exchange and we can therefore consider time-ordered sequences
\begin{equation}
    \{ t \}_{L} \equiv \{t_1, t_2, \dots, t_L \mid t_1 \ge t_2 \ge \cdots \ge t_L \} \, .
\end{equation}
The set of collective fields extends the bi-locals operational in the O($N$) symmetry case. The infinite sequence of multi-time collective variables will be shown to close under SD equations, giving a basis for the large-$N$ limit of these spin-chain models. We note that these S$_N$ invariants are due to the kinetic terms in the Lagrangian, as such their appearance is a universal feature for spin systems.

In the strong coupling limit ($1/J \rightarrow 0$), the potential term is dominant in the action~\eqref{eq:spin_XY4_Z_J}. Then, O($N$) symmetry emerges and the bi-local description applies. Since it is analogous to the bosonic SYK model, and has been shown by \cite{Baldwin:2019dki} that the replica non-diagonal configuration is of lower energy, we will consider the quenched averaging which involves $n$ replica fields. To compare with the well-known results in the SYK model, we will work in the Euclidean time $\tau = \ii t$. In the $1/J \rightarrow 0$ limit, we see that after rescaling
\begin{equation}
    \vec{S}_i \rightarrow J^{- \frac{1}{4}} \vec{S}_i \, ,
\end{equation}
the kinetic term in the action drops out, such that the replica representation of the partition function with the disordered average is
\begin{equation}
    \langle Z^n \rangle_{J} = \int \prod_{a=1}^{n} \prod_{i=1}^{N} \mathcal{D} S_{i,a}^+ \mathcal{D} S_{i,a}^- \, e^{- A[\vec{S}]} \, ,
\end{equation}
where $a$ is the replica index, and the Euclidean action is
\begin{equation} \label{eq:A_strong_coupling}
    A[\vec{S}] = -\frac{N}{4} \int d\tau_1 d\tau_2 
    \sum_{a,b=1}^{n}
    \left[
        \frac{1}{N}\sum_{i=1}^{N} \left(S_{i,a}^+(\tau_1) S_{i,b}^-(\tau_2) + S_{i,a}^-(\tau_1) S_{i,b}^+(\tau_2) \right)
    \right]^4 \, 
\end{equation}
with an emerging O($N$) symmetry $\vec{S}_i \rightarrow O_{i j} \vec{S}_j$. This allows a bi-local as the invariant collective field:
\begin{equation}
    \phi_{a b}(\tau_1, \tau_2) = \frac{1}{N} \sum_{i=1}^{N} \left[
        S_{i,a}^+(\tau_1) S_{i,b}^-(\tau_2) + S_{i,a}^-(\tau_1) S_{i,b}^+(\tau_2)
    \right] 
    \equiv
    \phi(X, Y)
    \, ,
\end{equation}
where we use the variable $X$ to package the time variable $t$ and the replica index $a$ \cite{Jevicki:2016bwu}. The bi-local field is symmetric under the exchange $X \leftrightarrow Y$:
\begin{equation}
    \phi(X, Y) = \phi(Y, X) \, .
\end{equation}
By contrast, the bi-local field in the SYK model is anti-symmetric \cite{Jevicki:2016bwu}. Thus, the partition function can be written as
\begin{equation}
    \langle Z^n \rangle_{J} = \int \mathcal{D} \phi(X, Y) \, \mathcal{J}[\phi] \, e^{- A[\phi]} \, .
\end{equation}
The Jacobian $\mathcal{J}[\phi]$ is
\begin{align}
    \mathcal{J}[\phi] & = \int \prod_{a=1}^{n}\prod_{i=1}^{N} \mathcal{D} S_{i,a}^+ \mathcal{D} S_{i,a}^- \,
             \delta\bigg(\phi(X, Y) - \frac{1}{N} \sum_{i=1}^{N}\big(S_i^+(X) S_i^-(Y) + S_i^-(X) S_i^+(Y)\big) \bigg)\, .
\end{align}
A standard way to deal with this is to introduce an auxiliary field, integrate out $S^\pm_i$, and then eliminate the auxiliary field by solving the saddle-point equations. The end result is
\begin{align} \label{eq:Jacobian_sol}
    \mathcal{J}[\phi]  = e^{N \Tr \ln \phi} \, .
\end{align}
With this Jacobian, we have the collective action in the strong coupling limit
\begin{equation}
    A_{\col}[\phi] = - \ln \mathcal{J} + A[\phi] = - N \Tr \ln \phi - \frac{N}{4} \sum_{X, Y} [\phi(X, Y)]^4 \, .
\end{equation}
In contrast to the SYK model~\cite{Jevicki:2016bwu}, the coefficient in front of the Jacobian term is minus instead of plus. As mentioned before, for the low temperature (large $\beta$) limit, replica indices should be added with possible replica non-diagonal solutions~\cite{Baldwin:2019dki}.
The SD equation is $\delta A_{\col}[\phi]/\delta\phi=0$, giving the relation
\begin{equation} \label{eq:phi_background_strong_coupling}
    \sum_{Z} [\phi_0(X, Z)]^{3} \phi_0(Z, Y) = - \delta_{X, Y} \, .
\end{equation}

We now consider finite coupling with only the S$_N$ symmetry and show the explicit form of the associated collective and SD equations. The general collective scheme for specifying the Jacobian $\mathcal{J}[\Phi]$ applies~\cite{Jevicki:1980zg}. It represents a change to the invariants $\Phi(\{t\}_L, \{t^\prime\}_L )$ defined in \eqref{eq:S_N_invariants}, with $L = 1, 2, \dots$. On the right-hand side of \eqref{eq:S_N_invariants}, the sum over $i$ is analogous to the trace in the matrix model. This sequence of `single-trace' fields is analogous to the `loop' or `word' variables of matrix models. Hence the basic building blocks will be the `splitting' and `joining' of `single-trace' fields.  
The `splitting' operation is 
\begin{align}
    \omega(\{t\}_L, \{t^\prime\}_L) = & \int \sum_{i=1}^{N} \frac{\delta^2 \Phi(\{t\}_L, \{t^\prime\}_L)}{\delta S_i^+(t) \delta S_i^-(t)} d t
    \nonumber \\
    = & \frac{1}{N} \sum_{l, k = 1}^{L} \delta (t_l - t^\prime_k)
        \sum_{i=1}^{N} S_i^+(t_1) \dots S_i^+(t_{l-1}) S_i^+(t_{l+1}) \dots S_i^+(t_L) \nonumber \\
      & \qquad \qquad \qquad \qquad \times   S_i^-(t^\prime_1) \dots S_i^-(t^\prime_{k-1}) S_i^-(t^\prime_{k+1}) \dots S_i^-(t^\prime_L) \, , 
\end{align}
resulting in a sum of variables of length $L - 1$. By using $a_L$ and $b_{L-1}$ to mean $(\{t\}_L, \{t^{\prime}\}_L)$ and $(\{t\}_{L-1}, \{t^{\prime}\}_{L-1})$, this operation can be written schematically as 
\begin{equation}
    \label{eq:omega_S_N}
    \omega(a_L) = \sum_{b_{L-1}} \Delta(a_{L}; b_{L-1}) \Phi(b_{L-1}) \, .
\end{equation}
Note that the counterpart of this operation in the matrix model splits a loop into two loops.
The `joining' is 
\begin{align}
    \Omega(a_L, & b_K) = \int \sum_{i=1}^{N} \frac{\delta \Phi(a_L)}{\delta S_i^+(t)} \frac{\delta \Phi(b_K)}{\delta S_i^-(t)} d t \nonumber \\
    = & \frac{1}{N^2} \sum_{l=1}^{L} \sum_{k=1}^{K} \, \delta(a_l - b^{\prime}_k) \sum_{i=1}^{N}
    S_i^+(a_1) \dots S_i^+(a_{l-1}) S_i^+(a_{l+1}) \dots S_i^+(a_L) S_i^-(a^\prime_1) \dots S_i^-(a^\prime_L) \nonumber \\
    & \qquad \qquad \qquad \qquad \times
    S_i^+(b_1) \dots S_i^+(b_K) S_i^-(b^\prime_1) \dots S_i^-(b^\prime_{k-1}) S_i^-(b^\prime_{k+1}) \dots S_i^-(b^\prime_K) \, ,
\end{align}
where the $l$-th and $k$-th spins are taken out of the sequence. This is then a linear combination of traces of length $L + K - 1$, or schematically
\begin{equation}
    \label{eq:Omega_S_N}
    \Omega(a_L, b_K) = \frac{1}{N} \sum_{c_{L + K - 1}} \Delta(a_L, b_K; c_{L + K - 1}) \Phi(c_{L + K - 1}) \, .
\end{equation}
Often, it is hard to obtain the Jacobian explicitly, while it is not hard to determine $\omega$ and $\Omega$ as illustrated above. Still, we can write the saddle-point equation explicitly without knowing the Jacobian~\cite{Jevicki:1980zg}:
\begin{equation} \label{eq:SD_collective}
    \sum_{b} \Omega(a, b) \frac{\delta A[\Phi]}{\delta \Phi(b)} - \omega(a) = 0 \, .
\end{equation}
This is the large-$N$ SD equation written explicitly in terms of the collective variables. As demonstrated in \cite{Jevicki:1980zg}, this general formula applies to the O($N$) vector model, U($N$) Yang-Mills gauge theory, etc. It applies to the large-$N$ spin systems as well, whose relevant collective variables are S$_N$ singlets, and one needs to substitute \eqref{eq:omega_S_N} and \eqref{eq:Omega_S_N} into this formula.
This set of equations represents a natural multi-time generalization of bi-local SD equations. It offers a possibility to search for more general ground state configuration of relevance at small temperatures. 

As a concrete example, we may apply these equations explicitly to the strong coupling limit. For simplicity let us assume that in this case we can have the replica-diagonal solutions such that we can ignore the replica indices. The action \eqref{eq:A_strong_coupling} can be written in terms of the $\Phi_1$:
\begin{equation}
    A[\Phi] = - \frac{N}{4} \int [\Phi_1(\tau, \tau') + \Phi_1(\tau', \tau)]^4 d \tau d \tau' \, .
\end{equation}
Since the action $A[\Phi]$ only depends on $\Phi_1$ in the strong coupling limit, we see that the SD equations \eqref{eq:SD_collective} reduce to
\begin{equation} \label{eq:SD_strong_couling_general}
    \int \Omega(\Phi_L, \Phi_1(\tau; \tau')) \frac{\delta A}{\delta \Phi_1(\tau; \tau')} d \tau d \tau'
    = \omega\left(\{\tau\}_L, \{\tau'\}_L\right) \, ,
\end{equation}
giving
\begin{align}
    - 2 \sum_{a=1}^L 
    \int [\phi(\tau_a, \tau) ]^{3}
    \Phi_L(
        \tau_1, \dots, \tau_{a-1}, \tau, \tau_{a+1}, \dots, \tau_L; \tau'_1, \dots, \tau'_L) 
    d \tau
    = \omega\left(\{\tau\}_L, \{\tau'\}_L\right) \, ,
\end{align}
where $\phi(\tau, \tau') = \Phi_1(\tau; \tau') + \Phi_1(\tau'; \tau)$. Explicitly, for $L=1,2$, we have
\begin{equation} \label{eq:Phi1_SD_strong_coupling}
    -2 \int 
    [\phi(\tau, \tau_1)]^3 \Phi_1(\tau; \tau_1') d \tau
    = \delta(\tau_1 - \tau'_1) \, ,
\end{equation}
\begin{align}
    -2 \int 
    \Big(
    [\phi(\tau, \tau_1)]^3 \Phi_2(\tau, \tau_2; \tau'_1, \tau'_2) + &
    [\phi(\tau, \tau_2)]^3 \Phi_2(\tau_1, \tau; \tau'_1, \tau'_2)
    \Big) 
    d \tau
    = \nonumber \\
    & \delta(\tau_1 - \tau'_1) \Phi_1(\tau_2; \tau'_2) +
    \delta(\tau_1 - \tau'_2) \Phi_1(\tau_2; \tau'_1) +
    (\tau_1 \leftrightarrow \tau_2) \, .
\end{align}
We see that the equation for $L=1$ \eqref{eq:Phi1_SD_strong_coupling} is consistent with the saddle point equation of the collective action \eqref{eq:phi_background_strong_coupling} we derived before. These equations have the recursive pattern that $\Phi_L$ is determined by the $\Phi_{L-1}$ and $\Phi_1$. 
Let $\Phi_L^{(0)}$ and $\phi_0$ be the solution of the $L=1$ part. Then, the following ansatz solves the above Schwinger-Dyson equations:
\begin{align}
    \Phi_L^{(0)} & (\tau_1, \dots, \tau_L; \tau_1', \dots, \tau_L')  \nonumber \\
    = & \frac{1}{2^L L!} \left[
        \int \phi_0(\tau, \tau') \sum_{i=1}^N \frac{\delta}{\delta S_i^+(\tau)} \frac{\delta}{\delta S_i^-(\tau')} \dd \tau \dd \tau'
    \right]^L
    \sum_{j=1}^N S_j^+(\tau_1) \dots S_j^+(\tau_L) S_j^-(\tau_1') \dots S_j^-(\tau_L') \\[5pt]
    = & \frac{1}{2^L} \sum_{\sigma \in \mathrm{S}_L} \phi_0(\tau_1, \tau'_{\sigma(1)}) \phi_0(\tau_2, \tau'_{\sigma(2)}) \dots \phi_0(\tau_L, \tau'_{\sigma(L)}) \, .
\end{align}
Thus, all multi-local fields are determined solely by $\Phi_1^{(0)}$, consistent with that in the strong coupling limit the only degree of freedom is the bi-local field.

\section{Toward quantum simulation}\label{sec:quantum-circuits}
We have already discussed in the introduction how the SYK4 model requires long chains of Pauli matrices when embedding the Majorana fermions on qubit degrees of freedom (e.g. using a Jordan-Wigner transformation). 
Those Pauli strings have a length that grows linearly with the size of the system $N_{\rm spin}$, making it prohibitively challenging to approach the many-spin ($N_{\rm spin}\to\infty$) limit.
On the other hand, the advantage of SpinXY4 over SYK4 is that each term in the Hamiltonian involves at most only four qubits, regardless of the size of the system. 
A review of the computational resources for the quantum digital simulation of the SYK4 model can be found in Ref.~\cite{Alvarez:2017abc,Babbush:2019abc} and a recent experimental trial for $N=6$ Majorana fermions on a superconducting qubit device has been reported in Ref.~\cite{Asaduzzaman:2023wtd}.

As an example of what building blocks are required for the digital quantum simulation of the dynamics of SpinXY4, we focus on a first-order Suzuki-Trotter decomposition and reduce the simulation to a product of 4-qubit unitary operations.
We can think of considering only spin operators acting on 4 different spins.
Practically, if $\hat{U}\equiv e^{-i J \delta t  \hat{\sigma}_{1,x}\hat{\sigma}_{2,x}\hat{\sigma}_{3,x}\hat{\sigma}_{4,x}}$ can be realized for $ J \delta t  \ll 1$, the Hamiltonian time evolution can be coded into a circuit using native single-qubit and two-qubit quantum gates.
We restrict to this exponential of a Pauli string because site indices can be handled by swapping qubit labels, and it is straightforward to replace $\hat{\sigma}_{j,x}$ with $\hat{\sigma}_{j,y}$ (or $\hat{\sigma}_{j,z}$) by a change of basis realized with single-qubit gates.
Let us note that having Pauli strings with several terms that are exponentiated is a very common occurrence in quantum chemistry applications~\cite{Whitfield_2011}, such as in the Unitary Coupled-Cluster ansatz, and there exist numerous techniques to synthesize the corresponding quantum circuits, such as those based on phase gadgets and \textsc{ZX}-calculus~\cite{Cowtan_2020}.

As an example, the unitary operator $\hat{U}$ above can be applied on 4 qubits using 6 CNOT gates in a staircase pattern, sandwiching a single-qubit Z rotation $R_{z}(\alpha) = e^{-\frac{1}{2} i \alpha \hat{\sigma}_{z}}$ with angle $\alpha = J \cdot \delta t$, while the Hadamard gate $H$ is used at the beginning and at the end of the circuit:
\[\Qcircuit @C=1em @R=.7em {
     & \gate{H} & \targ     & \targ     & \gate{R_{z}(J \cdot \delta t)} & \targ     & \targ     & \gate{H} & \qw\\
     & \gate{H} & \ctrl{-1} & \qw       &          \qw          & \qw       & \ctrl{-1} & \gate{H} & \qw\\
     & \gate{H} & \targ     & \ctrl{-2} &          \qw          & \ctrl{-2} & \targ     & \gate{H} & \qw\\
     & \gate{H} & \ctrl{-1} & \qw       &          \qw          & \qw       & \ctrl{-1} & \gate{H} & \qw\\
}\].

The Hamiltonian~\eqref{spin_Hamiltonian} can contain terms that are acting on 2 qubits, 3 qubits, or 4 qubits at most.
These terms will involve in general all qubits in the system, and all qubits will eventually be connected to all other qubits.
For the purpose of Trotterized digital quantum simulations, a system of qubits arranged with a local geometry will require a large number of \textsc{SWAP} gates to implement all the interactions. 
On the other hand, we can expect that trapped-ion devices, such as Quantinuum H-series systems~\cite{h-series-webpage}, can tame the non-local nature of the interaction.
In the case of the quantum charged-coupled device architecture of H-series~\cite{Pino:2020mku}, qubits are realized by ions that can physically move on the device, effectively implementing all-to-all connections with no additional gate overhead~\cite{shuttling-ions}.
One additional feature of the Quantinuum H-series systems is the native 2-qubit gate \textsc{ZZPhase}$(\alpha)$, which implements directly the operator $e^{-\frac{1}{2} i \alpha (\hat{\sigma}_{i,z} \otimes \hat{\sigma}_{j,z})}$ between any pair of qubits $i$ and $j$ with an infidelity that is proportional to the angle $\alpha$, and around $0.5 - 2.0 \times 10^{-3}$~\cite{Moses:2023ozv}.
Using such arbitrary-angle two-qubit gate we can express the circuit for $\hat{U}$ above with one less 2-qubit gate, replacing the Z rotation and the neighboring CNOT gates by a single \textsc{ZZPhase}:
\[\Qcircuit @C=1em @R=.7em {
     & \targ     & \gate{R_{z}(\alpha)}           & \targ     & \qw &                                          & & \multigate{2}{\textsc{ZZPhase}(\alpha)} & \qw \\
     &           &                                &           &     & \push{\rule{.3em}{0em}=\rule{.3em}{0em}} & &                                         &  \\
     & \ctrl{-2} & \        \qw                   & \ctrl{-2} & \qw &                                          & & \ghost{\textsc{ZZPhase}(\alpha)}        & \qw  \\
}\]
Overall, when taking into account the large number of terms in the Hamiltonian~\eqref{spin_Hamiltonian}, this results in a great reduction of the total circuit depth, making the circuit for the Trotterized simulation less susceptible to noise~\cite{Moses:2023ozv}.
In recent demonstrations of quantum algorithms on Quantinuum H-series devices, circuits with a number of 2-qubit gates between 600 and 1000 were run without significant loss of signal~\cite{yamamoto2023demonstrating} making use of tailored error detection techniques~\cite{self2022protecting}.
Moreover, in the application of quantum optimization algorithms, a recent paper has implemented circuits with $e^{-i \theta  \hat{\sigma}_{1,z}\hat{\sigma}_{2,z}\hat{\sigma}_{3,z}\hat{\sigma}_{4,z}}$ Hamiltonian terms on Quantinuum H-series devices with up to 1000 2-qubit gates~\cite{shaydulin2023evidence}, using an optimization algorithm to reduce the number of gates by arranging Hamiltonian terms.
The possibility of exploring the SpinXY4 variants described in Sec.~\ref{sec:variants}, such as introducing $\hat{\sigma}_{z}$ or reducing the number of terms to sparsify the interactions, using digital quantum simulations on real hardware is therefore a near-term challenge we would like to pursue in the future.
\section{Conclusion and discussion}
In this paper, we defined and studied the randomly coupled spin model (SpinXY4) by replacing Majorana fermions in the SYK model (SYK4) with Pauli spin operators. We found striking similarities between this model and the SYK model. We conclude that this is an interesting model of quantum chaos that can be simulated more easily on quantum computers. 

There are many directions to be explored. It would be nice if we could solve this model or some variants analytically. For the SYK model, the effective action in terms of bi-local fields provided us with a better understanding of the model itself and its relation to gravity. Hence, if we could do a similar analysis in terms of multi-local fields, we could understand if this model has a connection to gravity via holography. It would also be interesting to study various variants of the model including those suggested in Sec.~\ref{sec:variants}. We might be able to find an even simpler target for quantum simulation, or we might be able to find good models for holography or condensed matter physics. 

As a final remark, we point out the similarity between the SpinXY4 Hamiltonian and the interactions in the matrix model for quantum black hole (see Refs.~\cite{Gharibyan:2020bab,Maldacena:2023acv}). The matrix model contains several $N\times N$ matrices consisting of $N^2$ bosonic degrees of freedom. The interaction part of the Hamiltonian consists of O($N^4$) 4-local terms of these bosons. In the coordinate basis truncation, each bosonic operator can be written as a sum of $\hat{\sigma}_z$, and hence the entire interaction consists of the sum of 4-local interactions of $\hat{\sigma}_z$s. For this reason, the quantum simulation of SpinXY4 may be a good starting point for the simulation of the matrix model. Furthermore, the Yang-Mills theory can be embedded into the matrix model~\cite{Kaplan:2002wv,Buser:2020cvn}, and hence, the same technique can be used to study Yang-Mills theory, and probably, the standard model of particle physics. 
\begin{center}
\Large{
\textbf{Acknowledgements}}
\end{center}
\hspace{0.51cm}
We would like to thank Brian Swingle for stimulating discussions at various stages, including the initial suggestion of replacing fermions with hard-core bosons. We thank Brian Swingle and Michael Winer for sharing their study on SpinXY$q$ based on path integral. We also thank Tarek Anous, Budhaditya Bhattacharjee, Marcos Crichigno, Matthew DeCross, Michael Foss-Feig, Henry Lin, and Subir Sachdev.
Part of the computations in this work has been done using the facilities of the Supercomputer Center, the Institute for Solid State Physics, the University of Tokyo.
M.~H. and E.~R. thank the Royal Society International Exchanges award IEC/R3/213026. M.~H. also thanks the STFC grants ST/R003599/1 and ST/X000656/1.
A.~J. and X.~L. were supported by the U.S. Department of Energy under contract DE-SC0010010.
M.~T. was partially supported by the Japan Society for the Promotion of Science (JSPS) Grants-in-Aid for Scientific Research (KAKENHI) Grants No. JP20K03787 and JP21H05185.

\begin{center}
\Large{
\textbf{Data management}}
\end{center}
\hspace{0.51cm}
No additional research data beyond the data presented and cited in this work are needed to validate the research findings in this work. 
Simulation data will be available upon reasonable request.

\bibliographystyle{utphys}
\bibliography{XY4notes}

\end{document}